# Comparing Asset Pricing Models:
# Distance-based Metrics and Bayesian Interpretations

Zhongzhi (Lawrence) He[*]

This version: March 2018

## Abstract

In light of the power problems of statistical tests and undisciplined use of alpha-based statistics to compare models, this paper proposes a unified set of distance-based performance metrics, derived as the square root of the sum of squared alphas and squared standard errors. The Bayesian investor views model performance as the shortest distance between his dogmatic belief (model-implied distribution) and complete skepticism (data-based distribution) in the model, and favors models that produce low dispersion of alphas with high explanatory power. In this view, the momentum factor is a crucial addition to the five-factor model of Fama and French (2015), alleviating his prior concern of model mispricing by ±8% per annum. The distance metrics complement the frequentist p-values with a diagnostic tool to guard against bad models.

[*] Goodman School of Business, Brock University, 500 Glenridge Ave., St. Catharines, Ontario, Canada. Email: zhe@brocku.ca. I am grateful to Ken French who provided data for the Fama-French factors and the portfolio returns, to Lu Zhang who provided data for *q* factors, and to AQR who provided data for the monthly updated value factor. All errors are mine.

# Comparing Asset Pricing Models:
# Distance-based Metrics and Bayesian Interpretations

Multifactor asset-pricing models are designed to estimate expected returns for a cross section of LHS (left-hand-side) asset returns using a small set of RHS (right-hand-side) factors. A variety of multifactor models have been proposed, and models that produce low pricing errors (alphas) with high explanatory power (estimation precision) are deemed successful. However, the low-alpha and high-power criteria do not always reconcile, giving rise to what is known as the power problems that have long plagued asset pricing tests (Fama and French, 2012). As Barillas and Skanken (2017a) note, "a relatively large *p*-value (of statistical tests) may say more about imprecision in estimating a particular model's alphas than the adequacy of that model." The power problems afflict both the *t*-statistic for individual test and the *F*-statistic of *GRS* (Gibbons, Ross, and Shanken, 1989) for joint test of the zero-alpha restriction in time-series regressions.[1] Thus, comparing *t*- or *F*-statistics across different models can be problematic. Consequently, empirical studies rely on summary statistics of alphas jointly with the *GRS F*-statistic to judge model performance. Alpha-based statistics, especially the mean absolute alpha, are routinely reported as the main comparative results (e.g., Fama and French, 2015; 2016a, 2016b; Hou, Xue, and Zhang, 2015, 2016; Stambaugh and Yu, 2017, among many others);[2] the model with the lowest mean absolute alpha is deemed the best one. However, comparing models on alpha-based statistics is theoretically unfounded, nor does it consider the impact of differential explanatory power on model choice.[3] In particular, when the *GRS F*-statistic and mean absolute alpha produce contradicting model rankings (as commonly observed in the literature), which criterion should we follow and what are the causes of the ranking inconsistency? To what extent may the power problems (or ignoring the power) lead to a sub-optimal model choice? And how can we make sure that the lowest *t*- or *F*-statistic does not pick the model that blows up the alpha covariance matrix? In light of these open challenges, this paper theoretically motivates a unified set of distance-based performance metrics and provides coherent Bayesian interpretations in the context of model comparison.

---

[1] On the one hand, the powerful *GRS* test tends to over-reject models that produce economically insignificant pricing errors. For example, the three-factor model and the five-factor model of Fama and French (1993, 2015) produce an average pricing error of less than 0.10% per month for a wide range of test assets, but are still rejected by the *GRS* statistic that has power. On the other hand, the *GRS* test fails to reject models that produce large pricing errors with inflated alpha covariance matrix. For example, Fama and French (2012, 2016a) report that the *GRS* test cannot reject global models in pricing regional returns such as Japan due to lack of power. In a similar vein, Cochrane (2005) and De Moore, Dhaene, and Sercu (2015) raise the caution flags for blowing up the residual covariance matrix for a poor model to pass the statistical test.

[2] Other commonly reported comparative statistics include: the proportion of the magnitude or dispersion of unexplained average returns (Fama and French, 2015, 2016a, 2016b, 2017), the number of significant alphas by *t*-statistics (Hou, Xue, and Zhang, 2015, 2016), the number of rejections by the *GRS* test (Hou, Xue, and Zhang, 2015, 2016), the number of smallest magnitude of alpha (Stambaugh and Yu, 2017), the average absolute *t*-statistic (Stambaugh and Yu, 2017), among others.

[3] Barillas and Shanken (2017b) show examples that comparing models using alpha-based statistics may lead to inconsistent model rankings; however, they do not provide a formal analysis for this arguably ad hoc practice.

The distinction between the distance-based metrics and the ratio-based (*t*- and *F*-) statistics can be attributed to the fundamental difference between the frequentist interpretation and the Bayesian interpretation of model parameters. Specifically, a frequentist investor views alpha as a true but unknown parameter $\alpha$, and measurement error $\varepsilon$ prevents him from observing the true value of alpha, so the estimated alpha is $\hat{\alpha} = \alpha + \varepsilon$. To judge a model's performance, the frequentist investor measures the proportion of the dispersion of the alpha estimates attributable to estimation error, viewing a high ratio as good news: much of the dispersion of the alpha estimates is due to sampling error rather than to dispersion of the true alphas (Fama and French, 2016b, 2017). The power problem can thus arise as the statistical test may fail to reject models that produce large alphas but with inflated covariance matrix of alpha estimates. In contrast, a Bayesian investor views the sample as given to update his subjective belief about model mispricing, which is characterized by a posterior distribution of alpha $\alpha \sim N(\tilde{\alpha}, \tilde{V}_\alpha)$. The posterior estimates of mean $\tilde{\alpha}$ and variance $\tilde{V}_\alpha$ measure the expected value of mispricing and the mispricing uncertainty or estimation imprecision, respectively. This *probabilistic* view of the mispricing parameter makes it possible to define a unified set of performance metrics (namely, total distance, average distance, and marginal distance) between posterior distributions specified with distinct prior degrees of mispricing uncertainty $\sigma_\alpha$.[4] Intuitively, the Bayesian investor views model performance as the gap (in units of return) between his subjective belief and the objective reality. To evaluate performance, he rationally ranks all the models by their sizes of the gap, and picks the one with the smallest gap to be the best model.

In general, the distance metrics depend on specific values for the prior degree of mispricing uncertainty $\sigma_\alpha$. However, a frequentist equivalent version of the distance metrics does not, and it forms the upper bound of all distance measures for a given model in the sample. Specifically, when the Bayesian investor holds a dogmatic belief in the model *a priori* ($\sigma_\alpha = 0$), his posterior estimate of alpha shrinks to the model-implied value of zero with no uncertainty, i.e., a posterior distribution of zero mean and zero variance. At the other extreme, when the investor is completely skeptical about the model ($\sigma_\alpha = \infty$), the posterior estimates of alpha and mispricing uncertainty conform to their OLS sample estimates. The total distance (*TD*) metric is thus defined as the shortest distance between complete confidence (model-implied distribution) and complete skepticism (data-based distribution) in the model. Economically, *TD* is the minimum transport cost for the Bayesian investor to move the mass of the model-implied distribution to the data-based distribution (or in short, the minimum cost of holding his dogmatic belief in the model). *TD* is analogous to the *GRS* statistic in that both summarize the overall performance of the model in a single measure. However, unlike the ratio-based *F*-statistic (i.e., a relative statistical measure), the distance-based *TD* is derived as the square root of the *sum* of squared alphas and squared standard errors (i.e., an absolute cost measure). In contrast to the frequentist investor who prefers a low ratio of alpha estimates to sampling error, the Bayesian investor views both large dispersion of alphas and high mispricing uncertainty as bad news to the model, as they both contribute to enlarging the total distance. To the Bayesian investor, it is not the ratio, but the sheer size of the level and uncertainty of mispricing matters.

---

[4] Distance is not meaningfully defined under the frequentist view of alpha as a deterministic but unknown parameter.

The average distance (*AD*) is derived as the square root of the sum of mean squared pricing errors and mean squared standard errors. Compared to the total cost interpretation of *TD*, *AD* is interpreted as the minimum *average* cost of holding dogmatic belief in the model. Measured in units of average return in the root mean squared error (*RMSE*) sense, *AD* summarizes the contribution of pricing errors and that of standard errors for an average asset in the cross section; hence, it is used as the natural yardstick to compare models across different sets of assets. *AD* is akin to the mean absolute alpha in that they both measure the average performance of the model. However, compared to the mean-absolute-alpha metric that ignores model power and equally weights different magnitude of alphas, *AD* heavily penalizes extreme pricing errors and favors models that produce low alphas and high power with less extreme errors.

Finally, the marginal distance ($d_i$) is asset *i*'s contribution to the total distance, measuring the marginal cost of holding dogmatic belief in the model. $d_i$ to the Bayesian investor is analogous to *t*-statistic to the frequentist investor. To compare, *t*-statistic tends to identify significant (insignificant) pricing errors estimated with low (high) standard errors, i.e., the symptom of the power problems on individual assets. In contrast, $d_i$ is not a ratio, but is a cost measure by the square root of the sum of squared alpha and squared standard error of asset *i*. $d_i$ can be used to identify troublesome assets that contribute most to the total distance. Table 1 summarizes the comparative statics of the distance-based metrics and the *GRS* and mean absolute alpha statistics.

In the empirical part of the paper, the distance-based metrics are cast into a growing strand of literature on comparing models and choosing factors (e.g., Barillas and Shanken, 2017a, 2017b; Fama and French, 2017; Harvey and Liu, 2017). The Bayesian investor uses the average distance (*AD*) metric to rank the prominent models of different eras including recent competitors such as the Fama-French five factors (*FF5*), *FF5* plus the momentum factor (*FF6*), the *q*-factor model of Hou, Xue, and Zhang (2015) (*q*-factor), and the Bayes factor model of Barrilas and Shanken (2017a) (*BS*). The LHS assets cover a wide range of cross sections: four sets of 5×5 portfolios, three sets of 2×4×4 portfolios, plus 15 decile portfolios sorted by well-known anomalous variables. The distance-based model ranking results are compared with those based on the *GRS* and alpha-based statistics. We find that different performance metrics produce identical model ranking only for a small number of LHS assets. On the contrary, for the majority of cross sections, the distance metric produces different model rankings from those ranked by the *GRS* and alpha-based statistics that may themselves conflict with each other. From the viewpoint of the Bayesian investor, the ranking inconsistency is mainly attributed to two sources: 1) the power problems of the *GRS* test (to a larger extent), which tend to favor models that produce similar magnitude of alphas but are estimated less precisely; 2) the extreme-error problem of the mean absolute alpha (to a smaller extent), which treats all magnitude of alphas equally but ignores the detrimental effect of extreme pricing errors and/or standard errors on model performance. The two problems are best exemplified for the 25 Size-Momentum portfolios: both the *GRS* and mean absolute alpha pick *q*-factor the top model; however, *AD* ranks *FF6* the top model for its lowest dispersion of alphas and highest explanatory power.

In a nutshell, the Bayesian investor favors models that produce low dispersion of alphas with high estimation precision, but disproportionately dislikes extreme pricing errors and/or standard errors. By these criteria, *FF6*, *FF6-HML* (*HML* is dropped from *FF6*), and *q*-factor are the top three models, with the performance of the first two models indistinguishable from each other. Compared to the *GRS* and alpha-based statistics that often lead to counter-intuitive and contradicting ranking results, the *AD*-based ranking is consistently robust across the vast majority of cross sections. Much of the ranking discrepancy can be explained by the distinction between the frequentist view and the Bayesian view of the mispricing parameter. Specifically, using the *GRS* statistic, the frequentist investor typically picks *q*-factor the top model due to its lowest ratio of the dispersion of alpha estimates to sampling variation. In contrast, the Bayesian investor takes the sample as given to update his subjective beliefs, viewing model performance not by a ratio-based statistical measure but as a distance-based cost measure between two posterior distributions of the mispricing parameter. In this view, the Bayesian investor discovers that the top ranked

*q*-factor by the frequentist investor is not because of its better pricing ability, but rather due to its lower explanatory power. Instead, the Bayesian investor picks *FF6* to be the top model; intuitively, even at the same level of mispricing, *FF6* estimates alphas more precisely than *q*-factor.

We construct two distance measures to assess the economic magnitude of comparing models, or equivalently, to ascertain the economic significance of choosing factors. The first is the difference in the annualized total transport costs between two competing models. For example, adding the momentum factor (*UMD*) to *FF5* provides an annual saving of 4.5% to 7% for two broad cross sections. For non-nested models, *FF6* gains a competitive advantage over *q*-factor in the range of 2.5% to 3.2% annual savings of transport cost for the two cross sections. The second is a distance-equivalence measure for the Bayesian investor who moves away from his dogmatic belief in the model. Two competing models are said to be distance equivalent if they produce an identical average distance (*AD*) under different prior degrees of mispricing uncertainty. This measure entails an economic trade-off of model choice. To illustrate, *q*-factor at $\sigma_\alpha = 2\%$ is distance equivalent to a dogmatic belief in *FF6*. It means that choosing *FF6* over *q*-factor alleviates the Bayesian investor's prior concern of model mispricing by ±4% annually. Moreover, *FF5* at $\sigma_\alpha = 4\%$ is distance equivalent to an exact pricing *FF6*. This implies that the marginal value of the momentum factor (*UMD*) is to alleviate his prior concern of model mispricing by ±8% per annum. Therefore, the Bayesian investor regards *UMD* as a crucial addition to *FF5*, with its effect equally important as the joint effect of the profitability factor (*RMW*) and the investment factor (*CMA*) in describing average returns of broad cross sections. In contrast, the value factor (*HML*) is redundant, consistent with the frequentist finding (Fama and French, 2015). By the Occam razor's principle, the Bayesian investor therefore advocates a modified five-factor model with *HML* replaced by *UMD*.

This paper is closely related to Pastor and Stambaugh (2000) and Barrilas and Shanken (2017a). All three papers use a Bayesian framework to compare models and share the same prior specifications. Specifically, our paper builds on the Bayesian setting of Pastor and Stambaugh (2000). The probabilistic view of the mispricing parameter is required to define the distance between two multivariate normal distributions about alpha. Our paper differs from Pastor and Stambaugh (2000) in the objectives. The utility-based metrics of their paper are designed to examine the impact of varying degree of prior beliefs on portfolio choices, but not to "choose one pricing model over another". In comparison, our distance-based metrics can be used not only to measure the performance of a standalone model, but also to compare and rank models in both nested and non-nested settings. In this regard, our paper shares the same purpose as Barrilas and Shanken (2017a), but differs distinctly in the methodology. First, they use the Bayes factor to compute the posterior model probability in order to choose the best set of factors by statistical inferences. In comparison, our distance metrics are cost measures in units of return, carrying intuitive economic interpretations that are easy to communicate. Second, their paper belongs to the RHS approach based on spanning regressions of the candidate factors; ours is the LHS approach designed to pick the model that best describe the average returns of board cross sections.[5][6] Furthermore, the distance-based metrics more directly address the enduring challenges of the existing metrics. Total distance, average distance, and marginal distance are respectively comparable to the *GRS* statistic, alpha-based statistics, and *t*-statistic in the literature. Our Bayesian solution to the problems of the existing metrics is surprisingly simple. In its frequentist equivalent form, the OLS estimates of pricing errors and standard errors from time-series regressions are the only required inputs to construct the distance metrics, i.e., at no extra cost than constructing the *GRS* statistic. They only differ in the ways of assembling pricing errors and standard errors. Instead of taking the ratio, the Bayesian investor first adds up squared pricing errors and squared standard errors then takes the square root of the sum.

The rest of the paper is organized as follows. Section 1 presents the Bayesian framework and defines the quadratic Wasserstein distance in the optimal transport theory. Section 2 develops the metrics along with the Bayesian interpretations of total distance (*TD*), average distance (*AD*), and marginal distance ($d_i$). The properties of the distance metrics as compared to the *GRS* statistic, alpha-based statistics, and *t*-statistic are also analyzed. Section 3 presents data and time-series regressions. Sections 4 presents model ranking results for a wide range of LHS portfolios. Section 5 presents the Bayesian evaluation of the economic magnitude of comparing models and choosing factors. Section 6 concludes the paper with implications on the frequentist *p*-values.

---

[5] Fama and French (2017) classify choosing factors into the LHS approach and the RHS approach. The LHS approach judges competing models on the intercepts from times-series regressions to explain LHS returns. This is the dominant approach employed by a large number of studies, i.e., Fama and French (2015, 2016a, b), Hou, Xue, and Zhang (2015, 2016), Harvey and Liu (2017), just to list a few. Motivated by Fama (1998), the RHS approach uses spanning regressions to test for redundant factors and utilizes functions of the *GRS* statistic to make inferences. This approach has recently been advocated by Fama and French (2017) and Barrilas and Shanken (2017a, b), with its primary advantage being independent of LHS returns.

[6] Surprisingly, their six factors (*Mkt IA ROE SMB HML$^m$ UMD*) with the highest posterior probabilities by the Bayes factor measure do not perform well in describing LHS returns. As shown in Table 8, the six-factor model performs only better than *CAPM* and *FF3* in describing average returns for two board cross sections. The poor performance seems to be induced by the monthly updated value factor *HML$^m$*. Once *HML$^m$* is removed, the performance of the remaining five factors (*Mkt IA ROE SMB UMD*) significantly improve to the range between *q*-factor and *FF6*.

# 1. Bayesian Framework and Distance between Two Distributions

## 1.1 Bayesian Setting

The Bayesian setting is based on Pastor (2000) and Pastor and Stambaugh (2000). Assume an investor has $T$ observations on $n$ assets. Let $R$ denote the $T\times n$ matrix of asset returns in excess of risk free rate $r_f$, and $X = [l_T \quad F]$ denote the $T\times(k+1)$ matrix where the first column is a $T\times 1$ vector of ones and the remaining $k$ columns contain a $T\times k$ matrix of factor returns. Consider the multivariate regression of $R$ on $X$:

$$R = XB + U, \quad vec(U) \sim N(0, \Sigma \otimes I_T) \tag{1}$$

where $B = \begin{bmatrix} \alpha' \\ \beta' \end{bmatrix}$ is a $(k+1)\times n$ matrix where the first row is a $1\times n$ vector of alphas and the second row contains a $k\times n$ matrix of factor loadings, $l_T$ is an identity matrix of rank $T$. $vec(\cdot)$ denotes an operator that stacks the columns of a matrix into a vector, and $\otimes$ denotes the Kronecker product.

The rows of the disturbance matrix $U$ are assumed to be serially uncorrelated and homoscedastic with a $n\times n$ covariance matrix $\Sigma$. The setup of eq. (1) is the classic multivariate regression model considered by Zellner (1971) and applied by Pastor (2000) and Pastor and Stambaugh (2000) in traditional portfolio problems.

The investor's prior belief about $\alpha$ and $\beta$ is given by the following multivariate normal distribution:

$$p(B|\Sigma) \sim N\left(\begin{bmatrix} \alpha_0' \\ \beta_0' \end{bmatrix}, \begin{bmatrix} \sigma_\alpha^2(\Sigma/s^2) & 0 \\ 0 & \Psi \end{bmatrix}\right) \tag{2}$$

In general, $\alpha_0'$, which is the $1\times n$ prior mean of $\alpha$, can take any non-zero values to reflect the investor's proprietary views on the level of mispricing for the set of LHS assets. We set $\alpha_0' = 0$ to center the prior belief about the implication of an asset-pricing model, i.e., no mispricing given the true model. If the prior mean on alpha is centered on zero, then $\sigma_\alpha$ represents the investor's prior belief about the degree of mispricing or mispricing uncertainty. When $\sigma_\alpha = 0$, the investor has dogmatic belief in the model, so mispricing is completely ruled out. When $\sigma_\alpha = \infty$, the investor regards any level of mispricing equally likely, so relies on the data to detect the level of mispricing of a given model. Between these two extreme views, $\sigma_\alpha$ can take a wide range of values to express the investor's strong, modest, and weak beliefs in the model. Note that the prior uncertainty of $\alpha_0'$, $\sigma_\alpha^2(\Sigma/s^2)$, is proportional to the residual covariance matrix

Σ to reflect the fact that very large mispricing opportunities are improbable.[7] $s^2$ is a scalar whose value is set equal to the average of the diagonal elements of the sample estimate of Σ to make $\Sigma/s^2$ invariant to scaling. Finally, the prior distribution of factor loadings $\beta_0'$ is also centered about zero, with a diagonal covariance matrix Ψ whose elements take very large values to construct a non-informative prior on factor loadings.

In Appendix A, the prior distributions of $(B, \Sigma)$ are combined with the likelihood function for eq. (1) to derive the posterior estimates of regression parameters. Given the informative prior for $\alpha$ and non-informative priors for $\beta$ and $\Sigma$, the posterior estimates of regression parameters $(\tilde{B}, \tilde{\Sigma})$ have analytical solutions. In particular, for our choice of $\alpha_0 = 0$ and $\beta_0 = 0$, then the posterior estimates of $\alpha$ and $\beta$ have a simple and intuitive form:

$$\tilde{B} = \begin{bmatrix} \tilde{\alpha}' \\ \tilde{\beta}' \end{bmatrix} = (V_0^{-1} + X'X)^{-1} X'R \tag{3}$$

where $\tilde{\alpha}'$ is stored in the first row and $\tilde{\beta}'$ in the following $k$ rows of $\tilde{B}$; and $V_0^{-1} = \begin{bmatrix} s^2/\sigma_\alpha^2 & 0 \\ 0 & 0 \end{bmatrix}$ is a $(k+1)\times(k+1)$ matrix. The posterior variance of alpha $\tilde{V}_\alpha = Var[\alpha|R, F]$ is taken from the $(n, n)$ upper left block of the $n(k+1)\times n(k+1)$ matrix $\tilde{V} \otimes \tilde{\Sigma}$ defined in Appendix A.

The shrinkage effect for $\tilde{\alpha}$ is readily seen. For $\sigma_\alpha = \infty$ (complete disbelief in the model), $\tilde{B} = (X'X)^{-1}X'R$ reduces to the OLS estimates of $\alpha$ and $\beta$ based solely on the sample information. For $\sigma_\alpha = 0$ (dogmatic belief in the model), $\tilde{\beta}$ remains to be the OLS estimates, but $\tilde{\alpha}$ reduces to the theoretical value of zero. For any given value of $\sigma_\alpha$ between 0 and ∞, $\tilde{\alpha}$ is a weighted average of zero and the OLS estimates, with the respective weights determined by the relative confidence in the prior belief (captured by $V_0^{-1}$) and in the sample (captured by $X'X$).

For an asset pricing model, its mispricing is characterized by its posterior distribution of alpha $p(\alpha|R, F, \sigma_\alpha)$, with the value of $\sigma_\alpha$ specified from 0 to ∞ to reflect varying prior degrees of mispricing uncertainty. Thus, model performance is formulated as the problem of comparing two posterior distributions of alpha specified with distinct values of $\sigma_\alpha$. The optimal transport theory presented below provides the metrics that measure the shortest distance between two probability distributions.

---

[7] Pastor (2000) and Pastor and Stambaugh (2000) provide a detailed discussion for this prior specification. Barillas and Shanken (2017a) also use this prior and regard $k = \sigma_\alpha^2/s^2$ as the information ratio. He (2007) interprets $\sigma_\alpha$ as the active risk budget assigned to asset managers based on investment policies.

## 1.2 Optimal Transport Theory and Wasserstein Distance

The distance metrics are based on the optimal transport theory deeply rooted in mathematics (Villani, 2003, 2009) with rich applications in economics (Galichon, 2016) and econometrics (Galichon, 2017). The classic problem (Monge, 1781) is to find the shortest distance or the minimum cost to move the mass of one probability distribution to another. Following Villani (2009, Definition 6.1), we give the following definition of the Wasserstein distance between two probability distributions.[8]

**Definition**: Let $(S, d)$ be a Polish metric space. Assume that two probability measures $P_I$ and $P_{II}$ on $S$ are continuous and have finite moments of order $p \in [1, \infty)$. The Wasserstein distance between $P_I$ and $P_{II}$ is defined by

$$WD_p(P_I, P_{II}) = \left[\inf \int_S d^p(x,y) d\pi(x,y)\right]^{\frac{1}{p}}$$
$$= \inf\{[Ed(X,Y)^p]^{\frac{1}{p}}, \; law(X) = P_I \text{ and } law(Y) = P_{II}\} \tag{4}$$

The infimum is taken over all $\pi(x, y)$ in $\Pi(P_I, P_{II})$, which is the set of joint probability measures on random variables $X \times Y$ with marginals $P_I$ on $X$ and $P_{II}$ on $Y$.

Specifically, we are interested in $p = 2$ in this paper, and define the quadratic Wasserstein distance as:
$$WD_2(P_I, P_{II}) = \inf(E_\pi ||X - Y||^2)^{1/2} \tag{5}$$

The infimum is taken over all the transport plan $\pi(x, y)$ in $\Pi(P_I, P_{II})$, with the marginal distribution of $P_I$ on $X$ and $P_{II}$ on $Y$.

Given the definition of $WD_2$, the optimal transport literature proves the following theoretical results.

1) There exists a unique solution to the optimal transport problem of moving the mass of distribution $P_I$ to that of $P_{II}$. The one-to-one mapping is known as the optimal transport plan $y = T(x)$, where $x \sim P_I$ is mapped to $y \sim P_{II}$ via $T(x)$.

2) Under the optimal transport plan, random vectors $X \sim P_I$ and $Y \sim P_{II}$ are maximally correlated with each other.

Above are standard results in the optimal transport theory (Villani, 2003, 2009).[9] $WD_2$ has the economic interpretation as the minimum expected cost to transport the mass of distribution $P_I$ to $P_{II}$. In general, there exists no closed-form formula for $WD_2$ or $T(x)$ for general distributions. Fortunately, when $P_I$ and $P_{II}$ are Gaussian, closed-form formula for $WD_2$ and $T(x)$ can be derived, with the key results summarized in the following theorem.

---

[8] The Wasserstein distance is also known as the Monge-Kantorovich distance in the optimal transport literature (Villani 2009, Chapter 7). See also Villani (2009, Chapter 6) for a chronological review of this terminology.
[9] See Villani (2003, Theorem 2.12) for the proof of the existence and uniqueness of the solution in 1), with the maximum correlation property in 2) as a corollary. See also Galichon (2016, 2017) for the applications in economics and econometrics.

**Theorem**

Let $P_I$ and $P_{II}$ be two Gaussian measures on $R^n$ with finite second moments such that $P_I \sim N(\alpha_I, V_I)$ and $P_{II} \sim N(\alpha_{II}, V_{II})$, where $\alpha_I$ and $\alpha_{II}$ are two ($n \times 1$) vectors of mean, $V_I$ and $V_{II}$ are two ($n \times n$) symmetric, positive-definite covariance matrices, respectively.

The quadratic Wasserstein distance ($WD_2$) between $P_I$ and $P_{II}$ is given by

$$WD_2 = \sqrt{||\alpha_{II} - \alpha_I||^2 + ||V_{II} - V_I||} \qquad (6.1)$$

and

$$||V_{II} - V_I|| = Tr(V_I + V_{II} - 2(V_I^{1/2} V_{II} V_I^{1/2})^{1/2}) \qquad (6.2)$$

where $||\alpha_{II} - \alpha_I||$ is the Euclidean 2-norm of the mean difference vector; $||V_{II} - V_I||$ denotes the distance between the two covariance matrices; $Tr(\cdot)$ is the trace operator of a matrix; $V_I^{1/2}$ is the square root of the covariance matrix $V_I$ such that $V_I = V_I^{1/2} V_I^{1/2}$. For symmetric and positive-definite covariance matrix $V_I$, its ($n \times n$) square root matrix $V_I^{1/2}$ is unique, symmetric, and positive-definite.[10]

Appendix B outlines the proof with technical details.

## 2. Distance-based Metrics and Bayesian Interpretations

This section develops the Total Distance (*TD*), Average Distance (*AD*), and Marginal Distance ($d_i$), and analyzes their properties as compared to the *GRS* statistic, mean absolute pricing error (*MAE*), and *t*-statistic, respectively. Table 1 summarizes the comparative statics of the performance metrics in terms of their theoretical motivations, economic interpretations, and analytical properties.

[Insert Table 1 here]

### 2.1 Total Distance (*TD*) versus the *GRS* Statistic

To define the total distance within the Bayesian setting, the first two moments $(\alpha_I, V_I)$ of $P_I$ and $(\alpha_{II}, V_{II})$ of $P_{II}$ are respectively replaced by the model-generated posterior estimates of alpha and variance $(\tilde{\alpha}_I, \tilde{V}_{\alpha_I})$ and $(\tilde{\alpha}_{II}, \tilde{V}_{\alpha_{II}})$, where *I* and *II* represents two distinct prior degrees of mispricing uncertainty specified for a given pricing model. In particular, model *I* is specified with $\sigma_\alpha = 0$ (complete confidence in the model); under such a dogmatic belief, there is no uncertainty and the posterior estimate of alpha shrinks to theoretical value of zero, i.e., both $\tilde{\alpha}_I$ and $\tilde{V}_I$ are zero. On the other hand, model *II* is specified with $\sigma_\alpha = \infty$ (complete skepticism in the model), under which the posterior estimates $(\tilde{\alpha}_{II}, \tilde{V}_{\alpha_{II}})$ shrinks to their sample estimates based entirely on the data. Given such prior specifications, the distance metric $WD_2$ of eq. (6) reduces to:

$$TD = \sqrt{||\tilde{\alpha}_{II}||^2 + tr(\tilde{V}_{\alpha_{II}})} \tag{7.1}$$

---

[10] The symmetric and positive-definite $V_I^{1/2}$ is computed using the Schur algorithm (Deadman, Higham, and Ralha, 2013). Python library scipy.linalg.sqrtm() implements this algorithm.

The Bayesian interpretation of eq. (7.1) is summarized below:

*Proposition 1: Total Distance (TD)*

1) *TD is the shortest total distance between complete confidence (model-implied distribution) and complete skepticism (data-based distribution) in the model;*

2) *TD is the minimum total cost to transport the mass of the model-implied distribution to the data-based distribution; in short, TD is the minimum total cost of holding dogmatic belief in the model.*

Given the non-information priors of model *II*, as shown in Appendix, the posterior estimates $\tilde{\alpha}_{II}$ and $\tilde{V}_{\alpha_{II}}$ are identical to the maximum-likelihood estimate of alpha $\hat{\alpha}$ and its covariance matrix $\hat{V}_\alpha$, respectively. Hence, eq. (7.1) has its frequentist equivalent form as

$$TD = \sqrt{||\hat{\alpha}||^2 + tr(\hat{V}_\alpha)} \tag{7.2}$$

which is intuitively interpretable as follows.

*TD* is the square root of two sum-of-square components. $||\hat{\alpha}||^2$ or $||\tilde{\alpha}||^2$ (the sample and posterior estimates are used interchangeably given non-informative priors of model *II*), is the sum of squared alphas of the LHS assets; it measures the contribution of the dispersion of pricing errors to the total distance. The second component, $tr(\hat{V}_\alpha)$ or $tr(\tilde{V}_\alpha)$ is the sum of alpha variances of individual assets, and it measures the contribution of the estimation imprecision to the total distance. As such, *TD* is akin to the large-sample *GRS* statistic in the form of $\hat{\alpha}'\hat{V}_\alpha^{-1}\hat{\alpha}$. Both *TD* and *GRS* summarize the overall performance of a given model by a single measure that carries economic interpretations. Specifically, the core of the *GRS* statistic, $\hat{\alpha}'\hat{\Sigma}^{-1}\hat{\alpha}$, is the difference between the maximum squared Sharpe ratio of both the factors and assets *Sh(F, R)*² and that of the factors alone *Sh(F)*² (Barrilas and Shanken, 2017a, 2017b; Fama and French, 2017). In parallel, *TD* has a Bayesian interpretation as the minimum cost of holding dogmatic belief in the model.

However, despite their identical estimates of $\alpha$ and $V$, there is a key distinction between the frequentist interpretation of *GRS* and the Bayesian interpretation of *TD*, leading to distinct model ranking criteria.

To illustrate the difference, note that the two components in eq. (7.1) or eq. (7.2) are both sum-of-square terms, i.e., $||\tilde{\alpha}||^2 = \sum_{i=1}^n \tilde{\alpha}_i^2$ and $tr(\tilde{V}) = \sum_{i=1}^n \tilde{\sigma}_{\alpha,i}^2$, where $\tilde{\sigma}_{\alpha,i} = \tilde{V}_{\alpha(i,i)}^{1/2}$ is the posterior estimate of standard error of alpha for asset *i*. Dividing by the number of assets *n* matches *exactly* the values of $Aa_i^2$ (the average squared intercept) and $As^2(a_i)$ (the average squared standard error) as in Fama and French (2016b, 2017), respectively. The frequentist interpretation is best illustrated as follows:

*"… high values of $As^2(a_i)/Aa_i^2$ are good news: They say that much of the dispersion of the intercept estimates is due to sampling error rather than to dispersion of the true intercepts."* (Fama and French, 2016b, p.78 )

Thus, the frequentist interpretation examines the proportion of the dispersion of the alpha estimates due to sampling error, viewing the model that produces a higher ratio the better one. The *GRS* statistic is essentially in the same vein by taking the ratio of squared alphas to the covariance matrix of alpha estimates and favoring the model that produces lower *F*-statistic. The power problem thus originates out of this frequentist interpretation.[11] In contrast, the Bayesian interpretation views both large magnitude of alphas and large standard errors as bad news since they both contribute to enlarging the total distance. Thus, instead of weighting them against each other by taking the ratio, the distance metric takes the square root of the sum of the two squared errors; by doing so, *TD* is free of the power problem and penalizes models that estimates alphas less precisely. Furthermore, unlike Fama and French (2016b, 2017) that favors models with high $As^2(a_i)/Aa_i^2$, the Bayesian interpretation does not use this ratio as a ranking criterion; instead, a high ratio signals the symptom of the power problem. In particular, if this ratio is larger than one, it implies that the model is hard to reject due to lack of power, even though pricing errors produced by the model can be unlimited high.

### 2.2 Average Distance (*AD*) versus Alpha-based Statistics (*MAE*)

*TD* measures the *total* cost of transporting the mass of the model-implied distribution to the data-based distribution of mispricing. As a result, this total cost metric depends on the number of test assets, so may not be comparable across varying sizes of the investment universe. Below we propose an average cost metric akin to the alpha-based statistics (e.g., mean absolute alpha) in the literature, and can be used to compare models for differential size of cross sections.

---

[11] The power problems can be characterized as the scale problem inherent in all ratio-based metrics in the form of $\alpha'V_\alpha^{-1}\alpha$. Consider multiplying a scaler *k* to *V* and keeping $\alpha$ fixed. The ratio-based statistic $\alpha'(kV_\alpha)^{-1}\alpha$ can reject any model for *k*→0, and pass any model for *k*→∞. Intuitively, a good model requires both $\alpha$ and $V_\alpha$ to be small. The distance metric in the form of $\sqrt{||\alpha||^2 + tr(kV_\alpha)}$ reflects such a requirement thus effectively resolves the power problems.

For eq. (7.1), drop the subscript $II$ for ease of exposition and divide the two components by the number of assets $n$. Let's define

$$AD = \sqrt{RMSE^2(\tilde{\alpha}) + RMSE^2(\tilde{\sigma}_\alpha)} \qquad (8.1)$$

where

$$RMSE(\tilde{\alpha}) = \sqrt{\sum_{i=1}^{n} \tilde{\alpha}_i^2 / n} \qquad (8.2)$$

$$RMSE(\tilde{\sigma}_\alpha) = \sqrt{\sum_{i=1}^{n} \tilde{\sigma}_{\alpha,i}^2 / n} \qquad (8.3)$$

As seen, $RMSE(\tilde{\alpha})$ and $RMSE(\tilde{\sigma}_\alpha)$ are respectively the square *R*oot of *M*ean *S*quared pricing *E*rror and the square *R*oot of *M*ean *S*quared standard *E*rror. They measure the contribution of the average dispersion of pricing errors and the contribution of the average variance to the average distance, respectively. Similar to *Proposition 1*, the Bayesian interpretation of *AD* is

*Proposition 2: Average Distance (AD)*

1) *AD is the shortest average distance between complete confidence (model-implied distribution) and complete skepticism (data-based distribution) in the model;*

2) *AD is the minimum average cost to transport the mass of the model-implied distribution to the data-based distribution; or AD is the minimum average cost of holding dogmatic belief in the model.*

As an intuitive metric to compare models, *AD* is comparable to the widely used but undisciplined alpha-based statistics, in particular, the *M*ean *A*bsolute pricing *E*rror $MAE(\tilde{\alpha}) = \sum_{i=1}^{n} |\tilde{\alpha}_i|/n$. The inequality below gives the comparative statics of the performance metrics:

$$AD > RMSE(\tilde{\alpha}) \geq MAE(\tilde{\alpha}) \qquad (9)$$

The first strict inequality follows from the notion that any model is an incomplete description of cross-sectional expected returns (e.g., Fama and French, 2015) because $RMSE(\tilde{\sigma}_\alpha)$ is strictly positive, i.e., alpha estimates are imprecise. The second inequity is a standard statistical property of *RMSE*, and the equality holds if and only if the assets have identical magnitude of pricing errors. Furthermore, the wider the dispersion of pricing errors, the larger the gap between *RMSE* and *MAE*. To illustrate the distinction between the two criteria, suppose that two models are compared to explain returns on two assets: pricing errors produced by model *I* are 0.15% and 0.17% per month, and 0.05% and 0.25% per month by model *II*. If ranked by the low *MAE* criterion, model *II* is better with $MAE(\tilde{\alpha}_{II}) = 0.15\% < MAE(\tilde{\alpha}_I) = 0.16\%$. However, if ranked by the low *RMSE* criterion, model *I* is better with $RMSE(\tilde{\alpha}_I) = 0.1603\% < RMSE(\tilde{\alpha}_{II}) = 0.1803\%$. This example illustrates a notable property of the *RMSE* criterion: it gives relatively high weights to large pricing errors. As a result, it views models that produce extreme alphas particularly undesirable, so heavily penalizes these models by a large *RMSE*. For example, a model may produce low alphas for a wide range of assets, but performs very poorly on just a few assets (e.g., *FF5* for most anomalies but momentum). Then, the model's $RMSE(\tilde{\alpha})$ would be dominated by these few extreme alphas, and ranked lower than its competitor that produces a higher *MAE* but less extreme alphas. Such a model ranking inconsistency is termed the "extreme-error problem" in this paper.[12] Further note that the *RMSE* properties equally apply to $RMSE(\tilde{\sigma}_\alpha)$, which penalizes low-precision models, especially those that produce extreme standard errors even on a small number of assets. In contrast, $MAE(\tilde{\alpha}_I)$ does not consider estimation precision of the model, and it assigns equal weights to different sizes of pricing errors.

Finally, as the square root of the sum of $RMSE^2(\tilde{\alpha})$ and $RMSE^2(\tilde{\sigma}_\alpha)$, *AD* is the primary distance metric used to rank models. As illustrated above, *AD* tends to favor models that produce: 1) low pricing errors; 2) high estimation precision; and 3) less extreme pricing errors and standard errors.

### 2.3 Marginal Distance ($d_i$) versus *t*-statistic ($t_i$)

When an individual asset *i* is singled out of the total distance, we can define the marginal distance as

$$d_i = \sqrt{\tilde{\alpha}_i^2 + \tilde{\sigma}_{\alpha,i}^2} \tag{10}$$

Total distance is thus the square root of the sum of $d_i$ squared, i.e., $TD = \sqrt{\sum_{i=1}^n d_i^2}$. The Bayesian interpretation of $d_i$ is:

*Proposition 3: Marginal Distance ($d_i$)*

1) $d_i$ is the marginal contribution of asset *i* to the total distance;

2) $d_i$ is the marginal cost of holding dogmatic belief in the model.

---

[12] To discern the detrimental effect of the extreme-error problem, consider two models on five assets. Model *I* explains four assets perfectly (zero alphas) but produces an alpha of 50bp on the fifth asset, i.e., *MAE*($\alpha_I$) = 10bp and *RMSE*($\alpha_I$) = 22.26bp. Model *II* produces 20bp alphas on all five assets, so *MAE*($\alpha_{II}$) = *RMSE*($\alpha_{II}$) = 20bp. Thus, one extreme error destroys an otherwise perfect model.

$d_i$ is akin to the *t*-statistic ($t_i = \tilde{\alpha}_i / \tilde{\sigma}_{\alpha,i}$) to test the statistical significance of pricing error of asset *i*. $d_i$ and $t_i$ use the same input ($\tilde{\alpha}_i$ and $\tilde{\sigma}_{\alpha,i}$), but differ distinctly in the measure construction and interpretation. *t*-statistic is a ratio-based measure favoring models that produce smaller values: the estimate of pricing error is insignificant relative to its sampling error by the frequentist interpretation. However, an insignificant *t*-statistic may not be attributable to a small alpha estimate, but instead to an inflated standard error, causing the power problem on individual assets. In contrast, the Bayesian investor views the effect of asset *i* as its marginal contribution to the total distance of the cross section. Both large pricing errors and large standard errors are bad news to the asset. Thus, the marginal distance can help single out those individual assets that contribute most to the total distance.

To summarize, above are analytical illustrations of the distance-based metrics as compared to the traditional *GRS* and alpha-based statistics. Two problems, the power problem associated with the *GRS* statistic and the extreme-error problem associated with the mean absolute alpha, are identified through comparative statics and hypothetical examples. To what extent are these problems observed in data that lead to inconsistency in model ranking? How are the model ranking results by the distance-based metrics compared to those by the *GRS* and *MAE* metrics? These are empirical questions addressed in the remaining part of the paper.

## 3. Data and Time-Series Regressions

The data on factors include the Fama-French five factors (*MKT SMB HML RMW CMA*), the momentum factor (*UMD*) of Carhart (1997), the *ME*, *IA* and *ROE* from the *q*-factor model of Hou, Xue, and Zhang (2015), and the timely updated value factor *HML$^m$* proposed by Asness and Frazzini (2013) and used by Barrilas and Shanken (2017a). The sample period is from 1967:01 to 2016:12, totaling 600 observations. Table 2 shows the monthly mean return, standard deviation, and *t*-statistic for each factor in Panel A, and the correlation matrix of the ten factors in Panel B.

[Insert Table 2 here]

LHS portfolio returns cover a wide range of assets, including four sets of bivariate-sorted portfolios (25 *Size-B/M* portfolios, 25 *Size-OP* portfolios, 25 *Size-INV* portfolios, and 25 *Size-MOM* portfolios), three sets of three-way-sorted portfolios (32 *Size-B/M-INV* portfolios, 32 *Size-B/M-INV* portfolios, and 32 *Size-OP-INV* portfolios). Each set of LHS assets is examined individually and jointly as the first large cross section of augmented portfolios. In addition, following Fama and French (2016b) and Hou, Xue, and Zhang (2015), we also examine 15 sets of univariate-sorted decile portfolios that cover a board array of anomalies, most of which are not targeted by the respective factor model. These 15 decile portfolios are pooled into four groups: 1) FF-factor group containing 40 decile portfolios formed on market capitalization (*Size*), book-to-market (*B/M*), profitability (*OP*), and investment (*INV*); 2) valuation group containing 30 decile portfolios formed on earnings-to-price (*E/P*), cash flow-to-price (*CF/P*), and dividend yield (*D/P*); 3) prior return group containing 30 decile portfolios formed on momentum (*MOM*), short-term reversal (*STR*), and long-term reversal (*LTR*); and 4) other anomaly group containing 50 decile portfolios formed on accruals (*AC*), net share issues (*NI*), market beta (*beta*), variance (*VAR*), and residual variance (*RVAR*). We seek to explain excess returns on these four groups individually and jointly as the second large cross section of augmented portfolios. The two board cross sections are used to cross-verify model ranking consistency by various performance metrics. All the data on LHS assets are from the data library of Professor French.

Asset pricing models (factors) include *CAPM* (*MKT*), *FF3* (*MKT SMB HML*), *FF4* (*MKT SMB RMW UMD*), *FF5* (*MKT SMB HML RMW CMA*), *FF6* (*MKT SMB HML RMW CMA UMD*), *FF6-HML* (*MKT SMB RMW CMA UMD*), *q*-factor (*MKT ME IA ROE*), and *BS* (*MKT SMB HML$^m$ IA ROE UMD*). These models are compared and ranked for both individual sets of portfolios and the two board cross sections.

For each LHS portfolios *i*, its excess returns are explained by the factors as shown in the parentheses of each model. We run the following time-series regression using *FF6* as an example:

$$R_{it} - R_{ft} = \alpha_i + b_i MKT_t + s_i SMB_t + h_i HML_t + r_i RMW_t + c_i CMA_t + u_i UMD_t + \varepsilon_{it}, \quad (11)$$

where $R_{it}$ is month *t*'s return on portfolio *i* within a given set of the LHS assets, $t = 1967:01 – 2016:12$. From the time-series regressions for each set of LHS returns, the *GRS* statistic is computed as:

$$GRS = \frac{T-n-k}{n}[1 + Sh^2(F)]^{-1}\hat{\alpha}'\hat{\Sigma}^{-1}\hat{\alpha} \sim F_{n,T-n-k} \quad (12)$$

where *T* is the number of observations, *n* is the number of assets, *k* is the number of factors; $Sh^2(F)$ is the squared Sharpe ratio of the factors; $\hat{\Sigma} = \frac{1}{T}\sum_{t=1}^{T}\varepsilon_t\varepsilon_t'$ is the maximum likelihood estimate of the residual covariance matrix of LHS returns.

## 4. Model Performance and Model Ranking Results

The first three sub-sections present the model ranking results using different performance metrics, and explain ranking inconsistencies from the viewpoint of the Bayesian investor. Detailed descriptions are provided in Section 3.1 for the 5×5 bivariate-sorted portfolios, in particular the 25 *Size-MOM* portfolios that are most exemplary of the power and extreme-error problems. The economic magnitude of choosing one model over another is deferred to Section 3.4 and more to the Section 4.

### 4.1 Bivariate-sorted Portfolios

Table 3 show the distance-based metrics (the first five columns), compared with the existing metrics (the last four columns) for four sets of 5×5 bivariate-sorted portfolios, including 25 *Size-B/M* portfolios in Panel A, 25 *Size-OP* portfolios in Panel B, 25 *Size-INV* portfolios in Panel C, and 25 *Size-MOM* portfolios in Panel D. In the first two columns, *TD* is the total, and *AD* is the average cost of holding dogmatic belief in a given asset-pricing model. $RMSE(\tilde{\alpha})$ and $RMSE(\tilde{\sigma}_\alpha)$ in the next two columns give the component breakdown of *AD*, and are respectively the square root of mean squared posterior estimates of pricing errors ($\tilde{\alpha}$) and standard errors ($\tilde{\sigma}_\alpha$) generated by the data-based model. The 5$^{\text{th}}$ column $A|\tilde{\sigma}_\alpha^2|/A|\tilde{\alpha}^2|$ measures the contribution of the mispricing uncertainty to the distance relative to that of pricing errors. This ratio is identical to $As^2(a_i)/Aa_i^2$ proposed by Fama and French (2016b), but with a different use and interpretation. It is not used to rank models but to compare the power of models, i.e., the higher the ratio, the more imprecisely the model estimates alphas. In the last four columns, the *GRS* statistic, mean absolute alpha $A|\tilde{\alpha}|$, and $R^2$ are standard performance metrics in the literature (e.g., Hou, Xue, and Zhang, 2015). $A|\tilde{\alpha}|/A|\tilde{r}|$ is the ratio of mean absolute alpha to the mean absolute value of portfolio *i*'s mean return deviation from its cross-sectional average ($\tilde{r}_i = \bar{R}_i - \bar{R}$). Suggested by Fama and French (2015), $A|\tilde{\alpha}|/A|\tilde{r}|$ supplements the mean absolute alpha and measures the proportion of average returns left unexplained by a model.

[Insert Table 3 here]

For the 25 *Size-B/M* sorted portfolios in Panel A, the shortest distance or the total cost to move the mass of the model-implied distribution (dogmatic belief in the model) to the data-based distribution (complete skepticism in the model) is 1.633% per month under *CAPM*. Such a cost reduces to about half to 0.822% per month under *FF3*; it gets even smaller under other models, and reaches the lowest value of 0.665% under *FF6*. The average cost *AD* is intuitive to rank models. It ranks *FF6* the top model whose *AD* is 0.133% per month. This is consistent with its ranking by other statistics, i.e., lowest mean absolute alpha $A|\tilde{\alpha}|$ = 0.091%, lowest proportion of unexplained returns $|A|\tilde{\alpha}|/A|\tilde{r}|$ = 51%, and smallest *GRS* statistic of 2.93 (despite rejection at *p*-value < 1%). The model that drops *HML* (*FF6-HML*) ranks a close second by all the metrics. Rankings of other models are generally consistent but with one noticeable exception between *FF3* and *q*-factor. While *FF3* is ranked higher by its slightly lower alpha statistics, both the *GRS* and *AD* metrics rank the *q*-factor higher. The finding that *FF3* produces a lower $A|\tilde{\alpha}|$ (0.108% vs 0.111%) but a higher $RMSE(\tilde{\alpha})$ (0.15% vs 0.136%) suggests that there exist extreme pricing errors under *FF3*. This is demonstrated in Panel A of Table 4, which shows posterior estimates of alpha ($\tilde{\alpha}$), standard error ($\tilde{\sigma}_\alpha$), *t*-statistic ($\tilde{\alpha}/\tilde{\sigma}_\alpha$), and marginal distance ($d_i = \sqrt{\tilde{\alpha}_i^2 + \tilde{\sigma}_{\alpha,i}^2}$) for the 25 *Size-B/M* portfolios. As seen, the smallest-lowest *B/M* portfolio under *FF3* has an enormous alpha of -0.52% per month, whose magnitude is more than twice of all other alphas. In contrast, even though alphas produced by the *q*-factor model are on average larger, its greatest value (0.26%) is only half size of the extreme one under *FF3*. This extreme alpha is so detrimental to *FF3* that even its higher estimation precision (i.e., smaller $RMSE(\tilde{\sigma}_\alpha)$: 0.068% vs 0.085%; higher $R^2$: 91% vs 88%) does not overturn its performance relative to *q*-factor. The marginal distance $d_i$ summarizes both the dispersion of alphas and standard errors in one measure. $d_i$ = 0.53% for the lethal asset under *FF3* is more than twice of all other assets. This is the real-world demonstration that it takes just one extreme error to destroy an otherwise better model. Finally, the *GRS* statistic also favors the *q*-factor primarily due to its lower power as indicated by $A|\tilde{\sigma}_\alpha^2|/A|\tilde{\alpha}^2|$ = 39% under *q*-factor vs 21% under *FF3*.

[Insert Table 4 here]

For the 25 *Size-OP* sorted portfolios in Panel B of Table 3, first notice a substantially smaller total and average transport costs for all the models. Indeed, this set of portfolios produces the lowest *TD* and *AD* among all the bivariate-sorted and three-way-sorted portfolios, consistent with the finding of Fama and French (2015) that they are the best priced portfolios. Second, *FF5*, *FF6*, *FF6-HML*, and *q*-factor models all produce very similar performance metrics. However, inconsistent ranking still exists: the distance metric *AD* picks *FF5* or *FF6* to be the best model, whereas *q*-factor is ranked the top model by the *GRS* (1.48, *p*-value = 6.2%) and mean absolute alpha ($A|\tilde{\alpha}|$ < 0.06%). The discrepancy is not due to the extreme-error problem above; on the contrary, $RMSE(\tilde{\alpha})$ = 0.07% under *q*-factor is the lowest among all the models, but instead due to its lower power detected by $A|\tilde{\sigma}_\alpha^2|/A|\tilde{\alpha}^2|$ = 119%. As described earlier, the frequentist interpretation (Fama and French, 2016b) views this larger-than-one ratio positively: the dispersion of the true alphas is lower than that of sampling error. However, by the Bayesian interpretation, mispricing uncertainty contributes more to the distance than pricing errors, signaling a lack of power of the model. To elaborate on the power effect, Panel B of Table 4 shows the alphas ($\tilde{\alpha}$), standard errors ($\tilde{\sigma}_\alpha$), *t*-statistics ($\tilde{\alpha}/\tilde{\sigma}_\alpha$), and marginal distance ($d_i$) under the *FF5* and *q*-factor models for each *Size-OP* portfolio. $\tilde{\sigma}_\alpha$ is about 2 basis points higher under *q*-factor than under *FF5*. As a result, none of the alphas is rejected by the *t*-statistic under *q*-factor. For the same reason of low power, *q*-factor produces the lowest *GRS* test statistic among all models. In comparison, the marginal distance $d_i$ helps identify the asset that contributes the most to the total distance, which is the second largest and least profitable portfolio having $d_i = 0.19\%$ under *q*-factor. The *t*-statistic for this portfolio is 1.61, i.e., insignificant by the frequentist interpretation. However, its non-rejection by *p*-value > 10% is not because of its small alpha (0.16% is the largest under both models) but due to its highest standard error (0.10% is the highest under both models). Overall, the distance metric *AD* incorporates model's estimation precision into a cost component $RMSE(\tilde{\sigma}_\alpha)$, causing the model of lower explanatory power to be ranked lower.

For 25 *Size-INV* sorted portfolios in Panel C of Table 3, all performance metrics, i.e., *AD*, *GRS* and alpha-based statistics produces consistent model ranking, with *FF6* being the top model and the one that drops *HML* being the close second. No further discussion is needed here.

The distance-based metrics manifest their full merits in the set of the 25 *Size-MOM* sorted portfolios in Panel D of Table 3. It is well known that the *CAPM*, *FF3*, and *FF5* models fail to explain the cross-sectional average returns of momentum portfolios. From the viewpoint of a Bayesian investor, holding dogmatic belief in *FF5* incurs an enormous cost of 1.74% per month, and the average cost *AD* is economically highly significant at 0.35% per month. Adding the momentum factor (*UMD*) to *FF5* cuts *AD* by more than half to 0.16% per month. As for model ranking, *q*-factor is picked to be the top model by both the *GRS* (2.77, *p*-value < 1%) and alpha-based criteria ($A|\tilde{\alpha}| = 0.113\%$ and $A|\tilde{\sigma}_\alpha^2|/A|\tilde{\alpha}^2| = 42\%$). However, the average distance *AD* ranks *FF6* the top model, *FF6-HML* the close second, *FF4* the third, and *q*-factor in the fourth place. The primary reason for this ranking discrepancy is that the alpha-based statistics do not consider estimation precision nor extreme pricing errors, and the power problem of the *GRS* statistic tends to favor models with higher sampling error. Both the power problem and the extreme-error problem are evident in this cross section. First, *FF6* and *q*-factor models generate the same level of mean absolute alpha $A|\tilde{\alpha}| = 0.113\%$ and the proportion of unexplained returns of 42%. However, *q*-factor produces more extreme alphas than *FF6*, as evident by $RMSE(\tilde{\alpha}) = 0.158\%$ under *q*-factor vs 0.145% under *FF6*. More importantly, *q*-factor estimates alphas less precisely, hence results in greater alpha uncertainty, i.e., $RMSE(\tilde{\sigma}_\alpha) = 0.104\%$ under *q*-factor vs 0.073% under *FF6*, $A|\tilde{\sigma}_\alpha^2|/A|\tilde{\alpha}^2| = 43\%$ under *q*-factor vs 25% under *FF6*. The discrepancy in estimation precision is also evident in the $R^2$ values of the two models, $R^2 = 85\%$ under *q*-factor and 92% under *FF6*. Thus, given the similar level of pricing errors but estimated at different degrees of precision, the *GRS* test exhibits the well-known power problem that tends to reject the more accurate model (*FF6*) more heavily than the less precise one (*q*-factor). In comparison, the distance metric addresses both the power and extreme-error problems by one single summary statistic *AD*, which favors *FF6* (0.162%) over *q*-factor (0.19%). To delve into further details, Panel C of Table 4 shows the posterior pricing errors ($\tilde{\alpha}$), standard errors ($\tilde{\sigma}_\alpha$), *t*-statistics ($\tilde{\alpha}/\tilde{\sigma}_\alpha$), and marginal distance ($d_i$) for each of the 25 portfolios. We observe extreme pricing errors for the top performers in the three smallest winner groups, which are 0.40%, 0.22%, 0.19% under *FF6* compared to 0.50%, 0.28%, and 0.23% under *q*-factor. The *RMSE* criterion penalizes the larger extreme values under *q*-factor more heavily than the relatively smaller values under *FF6*. Furthermore, greater standard errors under *q*-factor cause the three extreme pricing errors to be rejected less severely (*t* = 4.68, 2.99, 2.42) than under *FF6* (*t* = 4.78, 3.44, 2.90), despite their larger economic magnitude under *q*-factor. The power problem of this nature produces a smaller *GRS* statistic for *q*-factor (2.77) than for *FF6* (3.34), which is counter-intuitive. The marginal distance $d_i$ in Panel C of Table 4 helps identify troublesome assets, which contribute most to the total distance by the *RMSE* criterion. The top three ones $d_i$ = 0.51%, 0.29%, and 0.25% under *q*-factor are larger than $d_i$ = 0.41%, 0.25%, and 0.23% under *FF6*. Therefore, the Bayesian investor prefers *FF6* to *q*-factor.

### 4.2 Three-way-sorted Portfolios

For the 32 *Size-B/M-OP* portfolios in Panel A of Table 5, we find that different performance metrics result in slightly different model rankings. First, the *GRS* and alpha-based statistics imply conflicting ranking results. While *q*-factor is ranked the top model by the *GRS* (1.89, *p*-value < 1%), alpha-based statistics $A|\tilde{\alpha}|$ = 0.111% and $A|\tilde{\alpha}|/A|\tilde{r}|$ = 48% tie *FF5*, *FF6*, and *FF6-HML* to be the best model. Given that *q*-factor produces larger pricing errors in both magnitude ($A|\tilde{\alpha}|$ = 0.127%) and dispersion ($RMSE(\tilde{\alpha})$ = 0.164%), the lowest *GRS* statistic is attributed to its more dispersed mispricing covariance matrix ($RMSE(\tilde{\sigma}_\alpha)$ = 0.117%). In comparison, the distance metric *AD* considers both the level and the uncertainty of mispricing, and ranks the top three models to be *FF6* (0.173%), *FF5* (0.175%), and *FF6-HML* (0.18%).

[Insert Table 5 here]

For the 32 *Size-B/M-INV* sorted portfolios in Panel B of Table 5, all performance metrics rank *FF6* and *FF6-HML* the top two models. For the 32 *Size-OP-INV* sorted portfolios in Panel C, *FF6* is ranked the top model by both the distance metric *AD* (0.161%) and alpha-based statistics $|\tilde{\alpha}|$ = 0.11% and $A|\tilde{\alpha}|/A|\tilde{r}|$ = 50%. However, it is *q*-factor that produces the smallest *GRS* statistic (2.60, *p*-value < 1%). Once again, this is due to the power problem of the *GRS* test. Given about the same magnitude and dispersion of pricing errors under *FF6* ($|\tilde{\alpha}| = 0.110\%$ and $RMSE(\tilde{\alpha}) = 0.142\%$) and *q*-factor ($|\tilde{\alpha}| = 0.113\%$ and $RMSE(\tilde{\alpha}) = 0.144\%$), the *GRS* statistic favors the one that estimates alphas less precisely, i.e, $RMSE(\tilde{\sigma}_\alpha) = 0.076$ under *FF6* vs 0.086 under *q*-factor.

### 4.3 Univariate-sorted Decile Portfolios

For the four groups of 15 decile portfolios sorted by anomalous variables, the results on the distance-based metrics, *GRS* statistic and alpha-based statistics are displayed in Panel A, B, C, and D of Table 6, respectively.

[Insert Table 6 here]

For the FF factor group (10 *Size* + 10 *B/M* + 10 *OP* + 10 *INV*) in Panel A, the top three models ranked by the distance metric (*AD*) are *FF6* (0.091%), *FF6-HML* (0.093%) and *FF5* (0.097%). These three models are consistent with the ranking by the *GRS* statistic (1.26, 1.27, 1.35) and alpha-based statistics ($A|\tilde{\alpha}| = 0.053\%, 0.054\%, 0.061\%$). It is not surprising that *q*-factor is not in the top three list, as the LHS assets are finer sorts of each of the four variables of *FF5*. In contrast, the remaining three groups of LHS assets are not targeted by the Fama-French five-factor model nor the *q*-factor model, thus a fair comparison.

For the valuation group (10 *E/P* + 10 *CF/P* + 10 *D/P*) in Panel B, the top three models ranked by *AD* are *FF3* (0.098%), *FF4* (0.104%) and *FF6* (0.131%). Once again, these top three models are consistent with the ranking by both the *GRS* statistic (0.99, 1.11, 1.36) and alpha-based statistics ($A|\tilde{\alpha}| = 0.051\%$, 0.056%, 0.086%). Note that *FF3* is the best model to explain the cross-sectional expected returns formed on valuation related firm characteristics. This result highlights the early success of *FF3* based on valuation variables (Fama and French, 1996). It means that, unlike all the other groups, adding extra factors to *FF3* does not help describe the average returns of the valuation group of assets. Also note that this is the only cross section for which *FF6* is not ranked the top model.

Inconsistent model rankings between the distance-based metrics and the *GRS* and alpha-based statistics re-emerge for the prior return group (10 *MOM* + 10 *STR* + 10 *LTR*) in Panel C. According to the *AD* metric, the top three models are *FF6* (0.143%), *FF6-HML* (0.143%), and *q*-factor (0.152%, which is very close to *FF4*, 0.153%). However, both the *GRS* (1.48, *p*-value < 10%) and mean absolute alpha ($A|\tilde{\alpha}| = 0.097\%$) pick *q*-factor to be the top model, and *FF6-HML* (*GRS* = 1.69, $A|\tilde{\alpha}| = 0.098\%$) and *FF6* (*GRS* = 1.76, $A|\tilde{\alpha}| = 0.104\%$) to be the 2$^{nd}$ and 3$^{rd}$ models. The discrepancy is largely attributed to the power problem, and to the extreme-error problem to a lesser extent for describing the 10 momentum portfolios. To see the details, Table 7 shows the pricing errors ($\tilde{\alpha}$), standard errors ($\tilde{\sigma}_\alpha$), *t*-statistics ($\tilde{\alpha}/\tilde{\sigma}_\alpha$), and marginal distance ($d_i = \sqrt{\tilde{\alpha}_i^2 + \tilde{\sigma}_{\alpha,i}^2}$) along with all the performance metrics for the 10 *MOM* portfolios. First, the top panel shows the results for *FF5*. The marginal distance of the bottom loser and top winner is 0.81% and 0.57% per month, respectively. By the *RMSE* criterion, these two extreme values overwhelmingly contribute to the average distance of 0.36% per month, suggesting the vital importance of adding the momentum factor (*UMD*) to *FF5*. The middle and bottom panels compare the performance of *FF6-HML* and *q*-factor. Despite the lower level of pricing errors under *q*-factor ($A|\tilde{\alpha}| = 0.121\%$) than under *FF6-HML* ($A|\tilde{\alpha}| = 0.127\%$), pricing errors are more dispersed under *q*-factor ($RMSE(\tilde{\alpha}) = 0.137\%$) than under *FF6-HML* ($RMSE(\tilde{\alpha}) = 0.133\%$), i.e., evidence of the extreme-error problem of *q*-factor. Examining individual alphas on the table, we find that *q*-factor produces the largest mispricing of 0.24% on the top winner, and the second largest mispricing of -0.18% on the bottom loser. In contrast, the largest and second largest mispricing under *FF6-HML* are 0.18% and 0.16%, respectively. More importantly, alphas are estimated less precisely by *q*-factor, $RMSE(\tilde{\sigma}_\alpha) = 0.109\%$ vs 0.072% under *FF6-HML*. This power problem results in lower *t*-statistics for individual test (e.g., for the top winter, $t = 1.94$ under *q*-factor vs $t = 2.09$ under *FF6-HML*) and smaller *GRS F*-statistic for the joint test ($F = 2.23$ under *q*-factor vs $F = 2.85$ under *FF6-HML*). The marginal distance is a more sensible metric to identify the largest contributing assets: $d_i = 0.27\%$ for the top winner and 0.26% for the bottom loser under *q*-factor. These two assets have among the highest magnitude of pricing errors (0.24% for the winner, -0.18% for the loser) and standard errors (0.13% for the winner, 0.19% for the loser). However, their *t*-statistics (1.94 for the winner, -0.98 for the loser) regard them as less significant than the $7^{th}$ decile asset ($t = -2.25$) with both a smaller alpha (-0.16%) and a smaller standard error (0.07%).

[Insert Table 7 here]

For the other anomaly group (10 *AC* + 10 *NI* + 10 *beta* + 10 *VAR* + 10 *RVAR*) in Panel D, the top three models are *FF6-HML*, *FF6*, and *q-factor* by *AD* (0.157%, 0.159%, 0.168%) and alpha-based statistics ($A|\tilde{\alpha}| = 0.100\%, 0.102\%, 0.112\%$). However, the *GRS F*-statistic has an identical lowest value (2.89, $p < 1\%$) for *q*-factor and *FF6-HML*. This is due to the relatively lower power as evidenced by $RMSE(\tilde{\sigma}_\alpha) = 0.082\%$ under *q*-factor, compared to 0.077% under both *FF6* and *FF6-HML*.

**4.4 Two Broad Cross Sections and Economic Magnitude**

We create two large cross sections from the above portfolios: the first one by augmenting four sets of 25 bivariate-sorted portfolios and three sets of 32 three-way-sorted portfolios, totalling 196 LHS returns; and the second one by augmenting 15 sets of univariate-sorted decile portfolios, totalling 150 LHS returns.

We rely on these two broad cross sections to give more robust model rankings that tend to diversify away biases induced by specific sorting variables. Furthermore, pooling together multiple sets of anomaly-based assets allows us to assess the economic magnitude of the distance-based metrics.

[Insert Table 8 here]

In addition to the models examined above, an all-factor model dubbed *All* that includes all the 10 factors is examined. This all-factor portfolio reflects the ultimate effort to mix different combinations of factors.[13] Table 8 shows the results on performance metrics for the two broad cross sections in Panel A and B, respectively. For both cross sections, the top three ranked models are *FF6*, *FF6-HML*, and *q*-factor by the *AD* metric. In the first set of 196 portfolios, the average distance for the top three models is 0.145%, 0.149%, and 0.165% per month, respectively; in the second set of 150 portfolios, the average distance is 0.135%, 0.135%, and 0.152% per month, respectively. However, the *GRS* and alpha-based statistics give inconsistent rankings for both cross sections. As described earlier, this is due to the low power of the *GRS* test (e.g., for the *q*-factor in the first set) and the extreme-error problem of alpha-based statistics (e.g., *q*-factor vs *FF4* in the second set). Finally, note that the all-factor model does not rank high in both cross sections, suggesting that blindly including all known factors does not produce the best model.

So far we have used the *average* cost metric (*AD*) to rank models across different sets of LHS assets. To assess the economic magnitude of comparing models and choosing factors, we turn to the *total* cost metric (*TD*) in the 1$^{st}$ column. First, it is readily seen that holding a dogmatic belief in *CAMP* and *FF3* incurs significantly higher costs than in the other models. For example, the total annual saving from a dogmatic belief in *FF3* to that in *FF5* is around (3.37% - 2.62%) $\times$ 12 $\approx$ 9% in the first broad cross section, and (2.62% - 2.03%) $\times$ 12 $\approx$ 7% in the second set. From the view of choosing factors, it means that adding *RMW* and *CMA* to *FF3* creates an additional value of 7% ~ 9% per annum for these two broad cross sections. Second, adding the momentum factor (*UMD*) to *FF5* creates economic significant savings, i.e., (2.62% - 2.04%) $\times$ 12 $\approx$ 7% per annum in the first set, and (2.03% - 1.65%) $\times$ 12 $\approx$ 4.5% per annum in the second set. Third, besides comparing nested models, the *TD* metric can compare non-nested models as well. For example, the competitive advantage of *q*-factor over *FF5* can be gauged as providing an additional saving of (2.62% - 2.31) $\times$ 12 $\approx$ 3.8% per annum in the first set, and (2.03% -1.86%) $\times$ 12 $\approx$ 2.1% per annum in the second set. However, just by adding *UMD* to *FF5*, *FF6* regains its competitive advantage over *q*-factor by (2.31% - 2.04%) × 12 ≈ 3.2% per annum in the first set, and (1.86% - 1.65%) × 12 ≈ 2.5% per annum in the second set. Finally, there is overwhelming evidence that *HML* is redundant for all sets of LHS assets. The performance of *FF6-HML* is indistinguishable from *FF6* not only in these two broad cross sections, but also in each individual set of portfolios described earlier. Therefore, in consideration of parsimony, one may consider substituting *UMD* for *HML* as a modified version of the five-factor model (*MKT SMB RMW CMA UMD*).

---

[13] For example, factors in *FF5* and *q*-factor can be mixed to create a new model (*MKT SMB HML IA ROE UMD*). Numerous factor combinations are examined. However, none of these mixed-factor models consistently outperforms *FF6* or *FF6-HML*.

Of course, the economic magnitude illustrated above depends on the investment universe, i.e., the larger the number of poorly diversified assets, the higher the total transport cost. Note that the LHS assets considered in this paper are all well-diversified portfolios. In the asset management industry, the idiosyncratic risk of investable assets can be much substantial. Thus, the economic magnitude of comparing models can be significantly larger than the above estimated. In this view, the *TD* metric can be economically meaningful for a portfolio manager who knows about his investment universe, e.g., the number of investable assets, the risk-return characteristics of the assets, etc. The manager is concerned about which asset-pricing model should be used to estimate expected returns with relatively higher precision. The distance metrices provide an economically meaningful measure in such a setting.

## 5. Comparing Models and Choosing Factors: A Bayesian Evaluation

The previous section compares models based on the distance metrics between complete model-based distribution ($\sigma_\alpha = 0$) and complete data-based distribution ($\sigma_\alpha = \infty$). This is the upper bound for the distance metrics for a given model in the sample. In this section, we consider what happens in between when investors express varying prior degrees of mispricing in a model. As Pastor (2000) argues, it might be reasonable to assume that investors neither use the model as a dogma nor do they regard the model as completely worthless. Influenced by empirical results from asset pricing tests, an investor may hold a wide range of prior degrees of mispricing in different models. For example, the recent success of *FF5* to explain a broad array of other anomalies but momentum may lead investors to an increased level of confidence but are still skeptical about the model to a certain degree; in the meantime, such empirical evidence may induce investors to cast further doubt on the validity of *FF3*, and even more so on *CAPM*. Since any prior degree of mispricing in the model can be translated into a distance metric between two posterior distributions, we can find the corresponding degrees of mispricing for *FF5*, *FF3*, and *CAPM* so that the posterior distributions generated by the three models have the same distance to their respective data-based distribution.

When $\sigma_\alpha$ is set from small values (strong prior belief in the model) to very large ones (highly skeptical about the model), the posterior estimate of alpha ($\tilde{\alpha}$) moves away from the theory value of zero towards its OLS sample estimate ($\hat{\alpha}$). As such, as soon as one departs from his dogmatic belief and casts a certain degree of doubt on the validity of the model, the mispricing distribution under his skeptical view is monotonically "closer" (than that under his dogmatic belief) to the data-based mispricing distribution by the *TD* or *AD* metric; in other words, the transport cost of holding a skeptical view in a model is always lower than holding the dogmatic belief in the model (i.e., the upper bound of the distance). The property of monotonically decreasing cost allows us to develop a distance-equivalent measure to assess the economic significance of comparing models and choosing factors. Specifically, we take the *AD* metric from Table 8 as the comparison benchmark for a given model ($\sigma_\alpha = 0$), then find $\sigma_\alpha > 0$ for an alternative model such that the two models produce an identical *AD*. For example, between *FF3* and *FF5*, there exists a unique $\sigma_{\alpha,FF3} > 0$ such that *AD* under *FF3* exactly equals to *AD* under *FF5* with $\sigma_{\alpha,FF5} = 0$. Thus, $\sigma_{\alpha,FF3}$ represents the degree of mispricing in *FF3* relative to *FF5* that sets the two models distance equivalent. Intuitively, it measures the following trade-off of model choice between *FF3* and *FF5*: if one chooses *FF3* over *FF5*, he is willing to accept $\pm 2 \times \sigma_{\alpha,FF3}$ (95% confidence interval) size of mispricing of *FF3 a priori*. Or conversely, the choice of *FF5* over *FF3* alleviates his prior concern of mispricing by $\pm 2 \times \sigma_{\alpha,FF3}$ per year. Of course, the larger the value of $\sigma_\alpha$, the more the relief of mispricing concern if one chooses the benchmark model over the alternative model. In this sense, the magnitude of $\sigma_\alpha$ can be used to assess the economic significance of comparing models and choosing factors.

We follow Pastor and Stambaugh (2000, p.353) and use the average volatility of the LHS assets as a guideline to specify the range and values of the prior mispricing volatility. The range of $\sigma_\alpha$ is from 0% to 10%, and the values are chosen to be 2%, 4%, 6%, 8% and 10%. The annualized volatility of individual portfolios in the two broad cross sections described above ranges from 10% to 30%, with the average volatility around 20%. The sample sizes corresponding to $\sigma_\alpha = 2\%, 4\%, 6\%, 8\%,$ and 10% are about 100 years, 25 years, 11.11 years, 6.25 years, and 4 years, respectively. In addition, $\sigma_\alpha = 2\%, 4\%, 6\%, 8\%,$ and 10% corresponds to a 95% confidence interval of mispricing of ±4%, ±8%, ±12%, ±16%, and ±20% per annum, respectively. By these assessments, $\sigma_\alpha = 2\%$ represents a modest degree of prior mispricing uncertainty (Pastor and Stambaugh, 2000), whereas other values of $\sigma_\alpha > 2\%$ represent significant prior degrees of mispricing for the investment universe.

[Insert Table 9 here]

Table 9 shows the distance metric *AD*, along with its component breakdown and relative contribution, for all the models at varying specified values of $\sigma_\alpha$. The results are for the same investment universe, i.e., the first broad set of 196 bivariate- and three-way-sorted portfolios on the left, and the second broad set of 150 univariate-sorted portfolios on the right. The results of $\sigma_\alpha = 0$ in Panel A are reproduced from Table 8 as the metrics of the benchmark model for comparison. According to the *AD* metric for the two board cross sections (196 portfolios, 150 portfolios), the model ranking from top to bottom is *FF6* (0.145%, 0.135%), *FF6-HML* (0.149%, 0.135%), *q*-factor (0.165%, 0.152%), *FF4* (0.187%, 0.166%), *FF5* (0.185%, 0.157%), *BS* (0.199%, 0.180%), *FF3* (0.241%, 0.214%), and *CAPM* (0.345%, 0.237%). As described earlier, these *AD* metrics represent the minimum costs of holding dogmatic beliefs in the corresponding models. We next assess the magnitude of trade-off between the benchmark model and the alternative model. We ask the following question: if one chooses the alternative model over the benchmark model, how much is the prior degree of mispricing he has to accept? The converse is probably more intuitive: by choosing the benchmark model over the alternative model, how much is the relief of his prior concern of model mispricing?

When $\sigma_\alpha > 0$, the posterior distribution of mispricing centers around non-zero means with non-zero covariance matrix, i.e., $\alpha_I | R, \sigma_\alpha \sim N(\tilde{\alpha}_I, \tilde{V}_I)$, where $\tilde{\alpha}_I$ and $\tilde{V}_I$ are the posterior estimates of mean and variance of model *I*. The total distance (*TD*) between $N(\tilde{\alpha}_I, \tilde{V}_I)$ and the data-based distribution $\alpha_{II} | R, \sigma_\alpha = \propto \sim N(\tilde{\alpha}_{II}, \tilde{V}_{II})$ generated by model *II* is computed from eq. (6). Then, the average distance (*AD*), its two components $RMSE(\tilde{\alpha})$ and $RMSE(\tilde{\sigma}_\alpha)$, and their relative contribution are subsequently computed. In this general Bayesian form, *TD* and *AD* have the economic interpretation as the total and average minimum cost of holding a certain degree of skeptical view (as specified by the non-zero $\sigma_\alpha$) about the model.

### 5.1 Comparing Models

[Insert Figure 1 Here]

Figure 1.A and 1.B plot *AD* at the specified values of $\sigma_\alpha$ for both board cross sections A and B. Each line represents one model, which, from top to bottom, is *CAPM*, *FF3*, *BS*, *FF5*, *FF4*, *q*-factor, *FF6-HML*, *FF6*, respectively.[14] We observe that all the lines are monotonically decreasing in a non-linear and non-

---

[14] Notice that certain pairs of models: *FF6* and *FF6-HML* in both figures, *FF4* and *FF5* in Figure 1.A, *FF4* and *q*-factor in Figure 1.B, are indistinguishable from each other (i.e., *AD* difference less than 0.5 basis points) at all specified values of $\sigma_\alpha$. These model pairs are viewed *AD* identical and allow us to identify redundant factors.

parallel pattern, i.e., the spread between the top and bottom lines shrinks from $\sigma_\alpha = 0$ to $\sigma_\alpha = 10\%$. This pattern reflects the shrinkage effect of the posterior estimates: as the skepticism in the model grows to a complete disbelief, all the monotonically decreasing lines eventually converge to zero.

To compare models using the distance-equivalent measure, we first identify the *AD* of a benchmark model on the vertical axis, then look horizontally to find the value of $\sigma_\alpha$ at which the *AD* of the alternative model equals or overlaps with that of the benchmark model. The value of $\sigma_\alpha$ carries the Bayesian interpretation as the prior degree of mispricing one has to accept if the alternative model is chosen over the benchmark model. Of course, such a trade-off is economically meaningful only if the benchmark model is the better performing one. For example, there is no alternative model to trade off if one dogmatically believes in *CAPM*; such an investor is willing to bear the average minimum cost of around 0.35% (0.24%) per month for the first (second) cross section for holding such a dogmatic belief. Below we interpret the distance-equivalent model comparison results in Figure 1 jointly with the numeric results from Table 9.

The results of $\sigma_\alpha = 2\%$ are reported in Panel B of Table 9. At this modest prior degree of mispricing, the posterior distribution of alpha is moderately "closer" to the data-based distribution. There are groups of models whose *AD*s at this level of mispricing (Panel B) overlap with those values on the vertical axis (Panel A). The most evident one is between *q*-factor (*AD* = 0.149% and 0.136% for the 1$^{st}$ and the 2$^{nd}$ cross sections) and *FF6*. It means that a modest skeptical view of *q*-factor at $\sigma_\alpha = 2\%$ is distance equivalent to a dogmatic belief in *FF6*. For the Bayesian investor, if he chooses *q*-factor over *FF6*, the prior belief is that an annualized pricing error of ±4% is acceptable to him; or conversely, his choice of *FF6* over *q*-factor alleviates his prior concern of model mispricing by ±4% per year. There are no other consistently overlapping models for both cross sections. However, overlapping models exist for a single cross section: for the set of 150 portfolios, *FF5* (*AD* = 0.149% in Panel B) overlaps with *q*-factor (*AD* = 0.152% in Panel A); for the set of 196 portfolios, *FF4* (*AD* = 0.168% in Panel B) is close to overlap with *q*-factor (*AD* = 0.165% in Panel A). In short, at a prior degree of mispricing at 2%, these alternative models are distance equivalent to their respective benchmark models.

Panel C displays the results of $\sigma_\alpha = 4\%$, which represents a significant prior degree of mispricing of ±8%. At this level of mispricing, the most noteworthy finding is that *FF5* (*AD* = 0.143% in Panel C) overlaps with *FF6* (*AD* = 0.145% in Panel A) for the set of 196 portfolios. Also, for the set of 150 portfolios, the *AD* of *FF5* (0.122% in Panel C) is below that of *FF6* (0.135% in Panel A); the more precise overlapping point is around $\sigma_\alpha = 3.6\%$. The Bayesian interpretation of this result is that, for an investor to choose *FF5* over *FF6*, he must be willing to accept an annualized pricing error of ±7 ~ 8%; conversely, the choice of *FF6* over *FF5* alleviates the prior concern of mispricing by ±7 ~ 8%, which can be attributed to the marginal value of *UMD* on top of *FF5*. Also note that, at $\sigma_\alpha = 4\%$ prior degree of mispricing, the *ADs* of *BS*, *FF3* and *CAPM* are still above that of the exact pricing *FF6*, suggesting that higher degrees of mispricing are required for these models to be distance-equivalent to *FF6*. It is surprising that the Bayes factor model *BS* does not perform well in describing average returns of these cross sections.

At $\sigma_\alpha = 6\%$ prior degree of mispricing in Panel D, *BS* and *FF3* are distance equivalent to an exact pricing *FF6*. An investor who chooses *FF3* over *FF6* is prepared to accept an annualized pricing error of ±12%. Thus, the failure of *FF3* is demonstrated from the perspective of a Bayesian investor, i.e., the relief of his prior concern of *FF3* mispricing is too large to be ignored. As for *CAPM*, it is distance equivalent at the prior mispricing uncertainty of 8% (Panel E) for the set of 150 univariate-sorted portfolios. However, for the set of 196 portfolios, even the prior mispricing uncertainty of 10% (Panel F) is not enough. This is not surprising, as the bivariate- and three-way-sorted portfolios are particularly designed to capture the failure of the *CAPM*, i.e., the beta is flat (Fama and French, 1992).

So far we have assumed complete confidence in the benchmark model. We now relax this assumption to reflect the more realistic scenario that investors hold varying degrees of skeptical views on a variety of models. For example, given the comparative results from recent empirical studies (Fama and French, 2015, 2016b; Hou, Xue, and Zhang, 2015, 2016), one may form certain degrees of confidence (but still skeptical) in *FF5* and *q*-factor. To what extent are these models distance equivalent? A horizontal glance at Figure 1 or cross-panel comparison of Panels B-F in Table 9 provides the answer. For example, both Figure 1.A and Figure 1.B show that *FF6* at $\sigma_\alpha = 2\%$ is distance equivalent to *q*-factor at $\sigma_\alpha = 4\%$. This means that choosing *FF6* over *q*-factor alleviates the investor's prior concern of mispricing by *additional* ±4% per year. Furthermore, when investors are highly skeptical in all models, the shrinkage effect of the posterior estimates makes the distinction between models smaller. For example, at $\sigma_\alpha = 8\%$ and 10%, *FF5* and *q*-factor become indistinguishable from each other.

### 5.2 Choosing Factors: *UMD* versus *HML*

The Bayesian analysis offers further insights into the choice of factors in an asset pricing model. We are particularly interested in the value factor (*HML*) and the momentum factor (*UMD*), both of which have received tremendous research attention with regard to their effectiveness in asset pricing.

Consistent with the findings of existing studies (Fama and French, 2015; Hou, Xue, and Zhang, 2015), our empirical results indicate that *HML* is indeed redundant. This is most intuitively viewed from Figure 1.A and 1.B, where the *AD* curves of *FF6* and *FF6-HML* are indistinguishable from each other at all prior degrees of mispricing. Therefore, *HML* adds no incremental benefit nor does it harm the model's ability to describe cross-sectional expected returns.

However, including the momentum factor (*UMD*) is crucial to the success of all Fama-French models to describe large cross sections. The marginal effect of including *UMD* can be assessed by examining the distance-equivalent metric between *FF6* and *FF5*. As the Bayesian results indicate, for an investor to choose *FF5* over *FF6*, he must be willing to accept the level of pricing errors in the range of ±7 ~ 8% per annum. In other words, the marginal effect of *UMD* is to relieve the investor's prior concern of mispricing by such a magnitude, which may be too large to be ignored. Furthermore, Figure 1 also shows the performance of *FF5* is no better than *FF4* (or even slightly worse in the set of 150 portfolios). This implies that the marginal effect of *UMD* is as large as (or even larger than) the joint effect of the profitability factor (*RMW*) and the investment factor (*CMA*), for reasons given below.

The momentum factor exerts two effects on the description of cross-sectional average returns. First and foremost, adding *UMD* drastically reduces the severity of the extreme-error problem of *FF5*, which is known to explain all other anomalies reasonably well but fail badly to explain momentum. By the *RMSE* property, however, the distance metrics are primarily determined by large pricing errors and large standard errors because they are ones that incur the greatest transport costs. Extreme pricing errors in the magnitude of the top winner and bottom loser as shown in the top panel of Table 7 predominantly contribute to the total or average distance. As demonstrated earlier, it takes just a few badly priced assets to destroy an otherwise good model, even in large cross sections like the two board cross sections. This is especially true for *FF5*, and explains why *UMD* has at least the same effect as *RMW* and *CMA*. Second, empirical results in all the tables show that, even for anomalies other than momentum, *FF6* consistently outperforms *FF5* in all cross sections, suggesting that adding *UMD* provides incremental value to the description of average returns of broad cross sections.[15]

---

[15] Adding factors does not necessarily improve model performance. For example, the *All*-factor model in Table 8 does not outperform *FF6*. Another example is $HML^m$ in the *BS* model, removing this factor improves the model performance close to *FF6*.

To summarize, in light of parsimony and the overwhelming evidence on the redundancy of *HML* but the crucial role of *UMD*, one may consider substituting *HML* with *UMD* in the five-factor model. We have shown throughout the paper that the performance of such a new five-factor model is indistinguishable from the best performing *FF6*.

## 6. Conclusion

In asset pricing tests, a model with high explanatory power may have low *p*-values (thus rejected) even if it produces economically insignificant alphas; conversely, a model with an inflated residual covariance of LHS returns may produce high *p*-values (thus pass the test) even with economically substantial alphas (Cochrane, 2005; De Moore, Dhaene, and Sercu, 2015). This nature of the power problems of statistical tests has long been recognized since Fama and French (1993), but the problems still afflict asset pricing tests nowadays (e.g., Fama and French, 2012, 2015, 2016). The power problems make comparing *p*-values across different models particularly problematic and the results hard to interpret (Harvey, 2017). As a compromise, empirical studies rely on various alpha-based statistics (which ignore the power problems) jointly with the *p*-values of the *GRS* test. However, the undisciplined use of the *GRS F*-statistic and alpha-based statistics is prone to counter-intuitive and contradicting model rankings, as commonly observed in the literature and showcased numerously in this paper.

The power problems may originate from the frequentist interpretation of the mispricing parameter as a deterministic but unknown value. As such, test statistics are derived from sampling theories of alpha estimates, favoring models that produce a low ratio (small *F*-statistic) of the dispersion of alpha estimates to sampling error (Fama and French, 2016b, 2017). This paper presents a Bayesian approach to address the enduring challenges in the context of model comparison. Below highlighted are the Bayesian interpretations of model parameter, model performance, and model comparison (see also Table 1).

- The Bayesian investor views model mispricing as a posterior distribution to represent his subjective belief updated with sample information. This probabilistic view of alphas enables the distance metrics to be meaningfully defined between two posterior distributions specified with varying prior degrees of mispricing uncertainty.

- The Bayesian investor views model performance as the shortest distance between his complete belief (model-implied distribution) and complete skepticism (data-based distribution). The

- distance is not a ratio, but a cost measure in units of return, carrying an economic interpretation as the minimum cost of holding dogmatic belief in the model.

- The frequentist equivalent form of the distance metrics is derived as the square root of the sum of squared alphas and squared standard errors. This is the upper bound of all distance measures for a given model in the sample, and is an objective measure for everyone (which is a desired property of the frequentist approach). The more general Bayesian form of the distance metrics is subjective and requires a specific value for the prior degree of mispricing uncertainty.

- Model performance is summarized by a unified set of total distance (*TD*), average distance (*AD*), and marginal distance ($d_i$), which are analogous to the *GRS* statistic for joint test, mean absolute alpha for average performance, and *t*-statistic for individual test in the frequentist setting, respectively.

- The Bayesian investor favors models that produce low dispersion of alphas with high explanatory power, but views extreme alphas and extreme standard errors highly undesirable. The goodness of model is more determined by how well the model prices those troublesome assets than the mean-absolute-alpha metric that treats all sizes of alphas equally.

When the above Bayesian interpretations are cast into the context of model comparison, the key empirical finding is that the momentum factor (*UMD*) may be more important to the Fama-French five-factor model than it was thought to be. The crucial role of the momentum factor is not merely documented for its ability to explain momentum portfolios (which we know that the five-factor model fails). Instead, various models compete to describe average returns for two board cross sections consisting of a majority of other variable sorts for which the five-factor model is known to perform well, and the momentum portfolios are just a small set (i.e., 25 out of 196, 10 out of 150). Even for these broad cross sections, *UMD* is a crucial addition to the five-factor model: by the distance equivalent metric to assess its economic significance, *UMD* relieves the Bayesian investor's prior concern of model mispricing by about ±8% per annum. This economic magnitude is equally significant as the joint effect of the profitability factor (*RMW*) and the investment factor (*CMA*). With the addition of *UMD*, the six-factor model (or drops the redundant *HML* factor for parsimony) becomes the best performing model among all competitors. Compared to the existing *GRS* and alpha-based statistics, the distance metrics generate economically intuitive and consistently robust model rankings across a board array of cross sections. In particular, low-power models that inflate the residual covariance matrix rank lower by the distance metrics, even if they may produce slightly smaller alpha estimates. It is essentially the resolution of the power problems that makes *FF6* ranked the top model, which would otherwise be the *q*-factor model for numerous cross sections.

Finally, the distance-based metrics along with the Bayesian interpretations may complement to the frequentist approach still dominant in asset pricing tests and model comparison, in light of Harvey's (2017) critiques on *p*-values. To a frequentist researcher, the distance metrics supplement at least two pieces of evidence toward the proper use and correct interpretations of *p*-values. First, the distance metrics in units of returns provide a coherent measure of the size of the economic effect that is lacking in statistical measures. Model choice is not viewed as a binary decision of whether the *p*-value passes a specific threshold; instead, the distance is a smooth cost measure to inform the decision maker of model performance in an economically meaningful way. Second, the distance metrics can be used to guard against a bad model that may look good on its *p*-values. If a model is not rejected or ranked high, is it due to its low pricing errors or low explanatory power? The total distance or average distance provides a diagnostic tool for the overall or average performance of the model, and the marginal distance may further help identify those troublesome assets that contribute most to model performance. Therefore, following Harvey's (2017) critique on *p*-values and suggested methodology, the frequentist *p*-values and the Bayesian distance metrics can be jointly used in asset pricing tests and model comparison.

# Appendix A. Posterior Estimates of Model Parameters

Below are standard conjugate results for the multivariate regression model of eq. (1).

The likelihood function of the model is

$$p(R|B,\Sigma) \propto |\Sigma|^{-\frac{T}{2}} \exp\{-\frac{1}{2}tr(R-XB)'(R-XB)\Sigma^{-1}\} \propto |\Sigma|^{-\frac{T}{2}} \exp\{-\frac{1}{2}tr[S+(B-\hat{B})'X'X(B-\hat{B})]\Sigma^{-1}\}$$
(A.1)

where $S = (R - X\hat{B})'(R - X\hat{B})$, $\hat{B} = (X'X)^{-1}X'R$, and $tr(\cdot)$ is the trace operator.

The prior distribution of model parameters is $p(B,\Sigma) = p(B|\Sigma)p(\Sigma)$, where

$$p(B|\Sigma) \sim N(B_0, \Sigma \otimes V_0) \propto |\Sigma|^{-\frac{k+1}{2}} \exp\{-\frac{1}{2}tr(B-B_0)'V_0^{-1}(B-B_0)\Sigma^{-1}\}$$
(A.2)

where $V_0^{-1} = \begin{bmatrix} s^2/\sigma_\alpha^2 & 0 \\ 0 & 0 \end{bmatrix}$ is a $(k+1)\times(k+1)$ matrix whose (1,1) element is $s^2/\sigma_\alpha^2$ and all other elements are zero; and

$$p(\Sigma) \sim IW(H_0, v_0) \propto |\Sigma|^{-\frac{v_0+n+1}{2}} \exp\{-\frac{1}{2}trH_0\Sigma^{-1}\}$$
(A.3)

is an inverted-Wishart distribution with degree of freedom $v_0 = n + 2$, so that the scale matrix is $H_0 = E[\Sigma] = \frac{s^2(v_0-n-1)I_n}{v_0-n-1} = s^2 I_n$

Combining the prior distribution (A.2) and (A.3) with the likelihood function (A.1) gives the following posterior distribution:

$$p(B,\Sigma|R) = p(R|B,\Sigma)p(B|\Sigma)p(\Sigma) \propto |\Sigma|^{-\frac{T+k+1+v_0+n+1}{2}} \exp\{-\frac{1}{2}tr[H_0 + S + (B-\hat{B})'X'X(B-\hat{B}) + (B-B_0)'V_0^{-1}(B-B_0)]\Sigma^{-1}\}$$
(A.4)

Completing the squares on B and collecting the remaining terms in [·] yields

$$H_0 + S + (B-\hat{B})'X'X(B-\hat{B}) + (B-B_0)'V_0^{-1}(B-B_0) = (B-\tilde{B})'\tilde{V}^{-1}(B-\tilde{B}) + \tilde{H}$$
(A.5)

where

$$\tilde{B} = (V_0^{-1} + X'X)^{-1}(V_0^{-1}B_0 + X'R)$$
(A.6)

$$\tilde{V} = (V_0^{-1} + X'X)^{-1}$$
(A.7)

$$\tilde{H} = H_0 + B_0'V_0^{-1}B_0 + S + \hat{B}'X'X\hat{B} - \tilde{B}'\tilde{V}^{-1}\tilde{B}$$
(A.8)

The posterior distribution (A.4) can be separated into two known distributions:

$$p(B,\Sigma|R) = p(B|\Sigma,R) \times p(\Sigma|R) \propto$$
$$\underbrace{|\Sigma|^{-\frac{k+1}{2}} \exp\left\{-\frac{1}{2}tr(B-\tilde{B})'\tilde{V}^{-1}(B-\tilde{B})\Sigma^{-1}\right\}}_{N(\tilde{B},\Sigma \otimes \tilde{V})} \times \underbrace{|\Sigma|^{-\frac{T+v_0+n+1}{2}} \exp\{-\frac{1}{2}tr\tilde{H}\Sigma^{-1}\}}_{IW(\tilde{H},\tilde{v})}$$
(A.9)

In words, $p(B|\Sigma, R)$ is normally distributed with posterior mean $\tilde{B}$ and posterior variance $\Sigma \otimes \tilde{V}$, and $p(\Sigma|R)$ is inverted-Wishart distributed with degree of freedom $\tilde{v} = T + v_0$ and scale matrix $\tilde{H}$. Denote the posterior estimate of the residual covariance matrix by $\tilde{\Sigma}$, which, from the properties of the Wishart distribution (Zellner, 1971), is calculated as

$$\tilde{\Sigma} = E[\Sigma|R] = \frac{\tilde{H}}{\tilde{v}-n-1} = \frac{\tilde{H}}{T+1}$$
(A.10)

From (A.9) and (A.10), the posterior distribution of alpha is normal with its posterior mean $\tilde{\alpha}' = E[\alpha|R,F]$ taken from the first row of $\tilde{B}$ and its posterior variance $\tilde{V}_\alpha = Var[\alpha|R,F]$ taken from the $(n, n)$ upper left block of $\tilde{V} \otimes \tilde{\Sigma}$.

With dogmatic belief in an asset-pricing model, mispricing is ruled out by setting the prior alpha uncertainty $\sigma_\alpha = 0$, so that both $\tilde{\alpha}$ and $\tilde{V}_\alpha$ are zero, i.e.,

$$p(\alpha|R, F, \sigma_\alpha = 0) \sim N(0,0) \tag{A.11}$$

At the other end of the spectrum $\sigma_\alpha = \infty$ where investors are completely skeptical about the model, the posterior mean $\tilde{\alpha}$ and variance $\tilde{V}_\alpha$ of alpha conform to the sampling theory results, which are

$$p(\alpha|R, F, \sigma_\alpha = \infty) \sim N(\bar{R} - \hat{\beta}\bar{F}, (1 + \bar{F}'\hat{\Omega}\bar{F})\frac{\hat{\Sigma}}{T}) \tag{A.12}$$

where $\bar{R}$ is a $(n \times 1)$ vector of sample mean of LHS excess returns, $\bar{F}$ is a $(k \times 1)$ vector of sample mean of factor returns, $\hat{\beta} = (F'F)^{-1}F'R$ is the $(n \times k)$ matrix of OLS estimates of factor loadings, $\hat{\Omega}$ is the $(k \times k)$ sample covariance matrix of factor returns, and $\hat{\Sigma} = (H_0 + S)/(T + 1)$ is the $(n \times n)$ residual covariance matrix of LHS returns estimated from the sample that dominates its non-informative prior $H_0$.

## Appendix B. The Distance between Two Multivariate Normal Distributions

Given two normally distributed random vectors $X \sim N(\alpha_I, V_I)$ and $Y \sim N(\alpha_{II}, V_{II})$ in $R^n$, define the demeaned random vectors $\underline{X} = X - \alpha_I$ and $\underline{Y} = Y - \alpha_{II}$. Define the squared quadratic Wasserstein distance $WD_2^2 \equiv E[||Y - X||^2] = ||\alpha_{II} - \alpha_I||^2 + E[||\underline{Y} - \underline{X}||^2]$. For ease of exposition, denote $||V_{II} - V_I|| \equiv E[||\underline{Y} - \underline{X}||^2]$. It follows that

$$WD_2 = \sqrt{||\alpha_{II} - \alpha_I||^2 + ||V_{II} - V_I||} \tag{B.1}$$

Under Gaussian measures, what remains to show is

$$||V_{II} - V_I|| = \text{Tr}(V_I + V_{II} - 2(V_I^{1/2} V_{II} V_I^{1/2})^{1/2}) \tag{B.2}$$

For the augmented random vector $(\underline{X}, \underline{Y})$ in $R^{2n}$, denote its covariance matrix by

$$\Psi = \begin{bmatrix} V_I & C \\ C^T & V_{II} \end{bmatrix} \tag{B.3}$$

Then $||V_{II} - V_I|| = \text{Tr}(V_I + V_{II} - 2C)$, and the infimum of $||V_{II} - V_I||$ is to find $C = E[\underline{XY}']$ so that $\underline{X}$ and $\underline{Y}$ are maximally correlated subject to the constraint that $\Psi$ is a positive definite covariance matrix. Thus, the optimization problem becomes

$$\underbrace{\max}_{C} 2\text{Tr}(C) \tag{B.4}$$

s.t.

$$V_I - C V_{II}^{-1} C^T > 0 \tag{B.5}$$

where (B.5) is the Schur complement constraint.

The solution of (B.4) subject to (B.5) leads to (B.2). The detailed proof is given by Dowson and Landau (1982) and Givens and Shortt (1984), where $WD_2$ is also termed the Frechet distance.

(B.2) can also be derived from the optimal transport mapping (Knott and Smith, 1984; Olkin and Pukelsheim, 1982). To check that the optimal transport plan maps $N(\alpha_I, V_I)$ to $N(\alpha_{II}, V_{II})$, for the zero-mean random vector $\underline{X} \sim N(0, V_I)$, let $\underline{Y} = T_I \underline{X}$, where the optimal mapping matrix is given by

$$T_I = V_I^{-1/2} \left(V_I^{1/2} V_{II} V_I^{1/2}\right)^{1/2} V_I^{-1/2} \tag{B.6}$$

Given $\underline{Y} = T_I \underline{X}$ and (A.6), we have

$$E[\underline{YY}'] = T_I E[\underline{XX}'] T_I' = V_I^{-\frac{1}{2}} \left(V_I^{\frac{1}{2}} V_{II} V_I^{\frac{1}{2}}\right)^{\frac{1}{2}} V_I^{-\frac{1}{2}} V_I V_I^{-\frac{1}{2}} \left(V_I^{\frac{1}{2}} V_{II} V_I^{\frac{1}{2}}\right)^{\frac{1}{2}} V_I^{-\frac{1}{2}} = V_I^{-\frac{1}{2}} \left(V_I^{\frac{1}{2}} V_{II} V_I^{\frac{1}{2}}\right)^{\frac{1}{2}} \left(V_I^{\frac{1}{2}} V_{II} V_I^{\frac{1}{2}}\right)^{\frac{1}{2}} V_I^{-\frac{1}{2}} =$$

$$V_I^{-\frac{1}{2}} \left(V_I^{\frac{1}{2}} V_{II} V_I^{\frac{1}{2}}\right) V_I^{-\frac{1}{2}} = V_{II} \tag{B.7}$$

In the univariate case where $V_I = \sigma_I^2$ and $V_{II} = \sigma_{II}^2$ are scalers, the optimal mapping matrix simplifies to a scaler $T_I = \sigma_I^{-1}(\sigma_I \sigma_{II}^2 \sigma_I)^{1/2} \sigma_I^{-1} = \sigma_{II}/\sigma_I$. Then $\sigma_{II}^2 = (T_I \sigma_I)^2$ is easily verified.

To check that $T_I$ is indeed optimal, we have

$$E[||\underline{Y} - \underline{X}||^2] = E[||\underline{X}||^2] + E[||\underline{Y}||^2] - 2E[\langle \underline{X}, \underline{Y} \rangle] = Tr(V_I) + Tr(V_{II}) - 2E[\langle \underline{X}, T_I \underline{X} \rangle] = Tr(V_I) + Tr(V_{II}) - 2tr(V_I T_I) = Tr(V_I) + Tr(V_{II}) - 2tr\left(\left(V_I^{\frac{1}{2}} V_{II} V_I^{\frac{1}{2}}\right)^{\frac{1}{2}}\right) = Tr(V_I + V_{II} - 2\left(V_I^{\frac{1}{2}} V_{II} V_I^{\frac{1}{2}}\right)^{\frac{1}{2}})$$

(B.8)

The second last equality is by the cyclic property of the trace operator.

The converse optimal transport mapping can also be derived. Let $\underline{X} = T_{II} \underline{Y}$, where the optimal mapping matrix is given by $T_{II} = V_{II}^{-1/2} \left(V_{II}^{1/2} V_I V_{II}^{1/2}\right)^{1/2} V_{II}^{-1/2}$. It is easy to verify that $T_{II} = T_I^{-1}$.

Figure 1. Average Distance under Varying Prior Degrees of Mispricing Uncertainty

This figure depicts the average distance (*AD*) created by a variety of models at varying prior degrees of mispricing uncertainty $\sigma_\alpha$ = 0, 2%, 4%, 6%, 8%, 10%. Asset pricing models (factors) include: *CAPM* (*MKT*), *FF3* (*MKT SMB HML*), *FF4* (*MKT SMB RMW UMD*), *FF5* (*MKT SMB HML RMW CMA*), *FF6* (*MKT SMB HML RMW CMA UMD*), *FF6-HML* (*MKT SMB RMW CMA UMD*), *q*-factor (*MKT ME IA ROE*), and *BS* (*MKT SMB HML$^m$ IA ROE UMD*). Figure 1.A is for an augmented cross section of four sets of 5×5 bivariate-sorted portfolios and three sets of 2×4×4 three-way sorted portfolios. Figure 1.B is for an augmented cross section of 15 decile portfolios sorted by anomaly variables. Sample period is 1967:01 – 2016:12, 600 months.

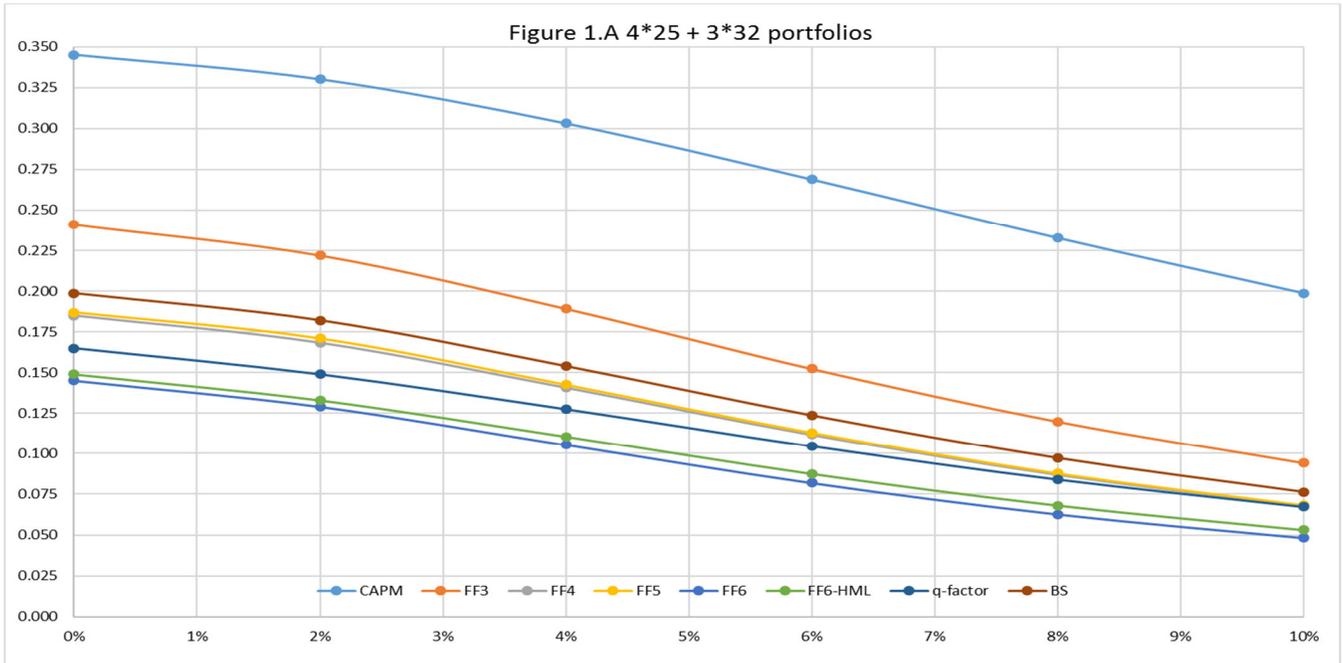

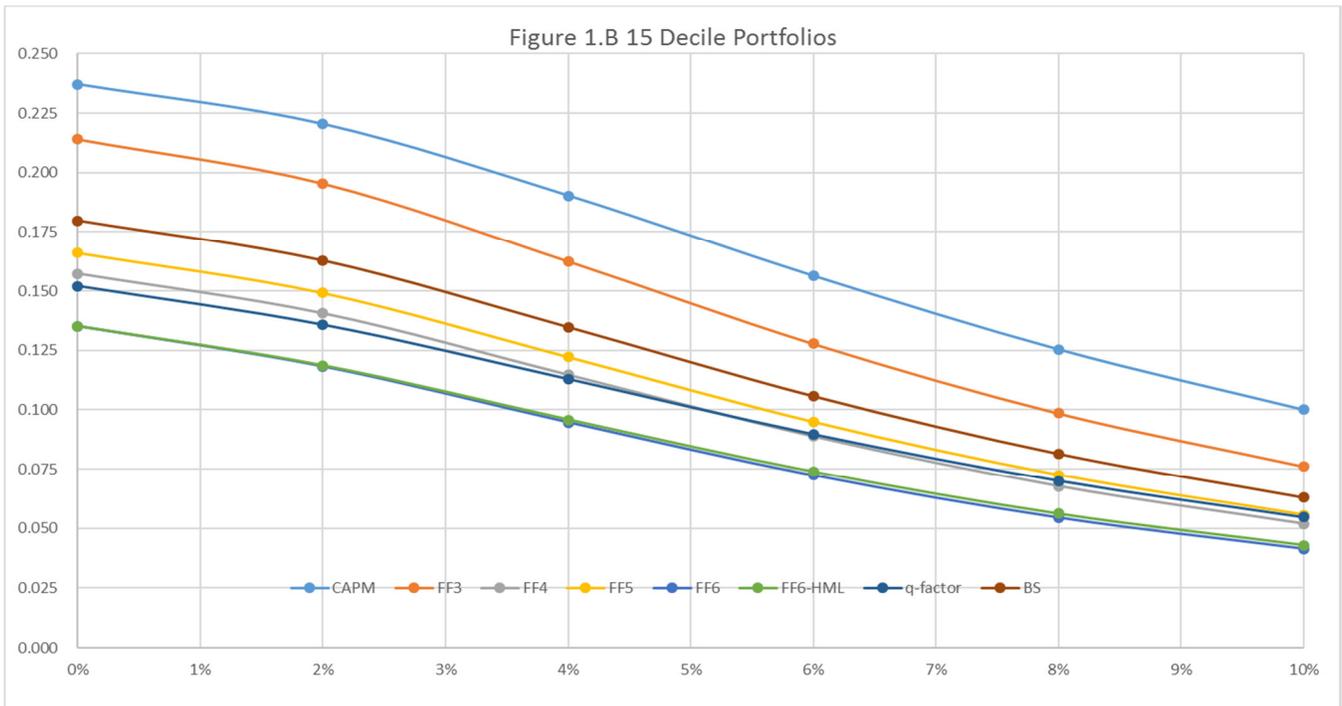

Table 1. Performance Metric Comparison: Distance Metrics, *GRS* and Mean Absolute Pricing Error (*MAE*)

*TD*, *AD*, and $d_i$ are total distance, average distance, and marginal distance, respectively. $\tilde{\alpha}_i$ and $\tilde{\sigma}_{\alpha,i}$ are the posterior estimates of pricing error and standard error of asset *i* with non-informative priors. Hence, $\tilde{\alpha}_i$ and $\tilde{\sigma}_{\alpha,i}$ are identical to their sample estimates $\hat{\alpha}_i$ and $\hat{\sigma}_{\alpha,i}$. $\hat{V}_\alpha$ is the covariance matrix of alpha estimates. $\hat{\Sigma}$ is the maximum likelihood estimate of the residual covariance matrix of test-asset returns. $Sh^2(F)$ is the squared Sharpe ratio of factors. *T*, *n*, *k* are respectively the number of observations, the number of assets, and the number of factors.

| Criteria/Metrics | *TD* and *AD* | *GRS* | *MAE* |
|---|---|---|---|
| Definition | $TD = \sqrt{\sum_{i=1}^n \tilde{\alpha}_i^2 + \tilde{\sigma}_{\alpha,i}^2}$<br>$AD = \sqrt{\sum_{i=1}^n (\tilde{\alpha}_i^2 + \tilde{\sigma}_{\alpha,i}^2)/n}$ | $GRS = \hat{\alpha}'\hat{V}_\alpha^{-1}\hat{\alpha}$, where $\hat{V}_\alpha^{-1} = [\frac{T-n-k}{n}\frac{1}{1+Sh^2(F)}]\hat{\Sigma}^{-1}$ | $MAE = \sum_{i=1}^n |\tilde{\alpha}_i|/n$ |
| Marginal Contribution of Asset *i* | $d_i = \sqrt{\tilde{\alpha}_i^2 + \tilde{\sigma}_{\alpha,i}^2}$ | $t_i = \hat{\alpha}_i/\hat{\sigma}_{\alpha,i}$ | $|\tilde{\alpha}_i|$ |
| Measurement Unit | Percentage return | *F*-statistic or *t*-statistic | Percentage return |
| Mispricing Parameter | Bayesian View:<br>Alpha is random, data is given to update posterior distribution of alpha; Performance is distance-based, small alphas and high estimation precision are preferred (regardless of the ratio) | Frequentist View:<br>Alpha is fixed, data is random; rely on sampling theory to derive test statistics; Performance is ratio-based, a low ratio of alpha estimates to sampling error is preferred | Simple statistical artifact regardless of the view |
| Theoretical Motivation | Bayesian method (Pastor, 2000; Pastor and Stambaugh, 2000); Optimal Transport Theory (Villani, 2003) | Sampling theory of multivariate statistics (Gibbons, Ross, and Shanken, 1989) | Ad hoc statistical measure |
| Economic Interpretation | The shortest distance to transport the mass of a pure model-implied distribution of mispricing to the pure data-based distribution of mispricing; The minimum cost of holding dogmatic belief in the model | $\hat{\alpha}'\hat{\Sigma}^{-1}\hat{\alpha}$ is the difference between the maximum squared Sharpe ratio of both the factors and assets and that of the factors alone; A statistical test if the factors span the mean-variance-efficient tangency portfolio | No theory or explanation to justify the use of absolute alphas |
| Pricing Errors (Alphas) | $RMSE(\tilde{\alpha}) = \sqrt{\sum_{i=1}^n \tilde{\alpha}_i^2/n}$ regards large pricing errors highly undesirable, and heavily penalizes models that produce extreme alphas | Squared alphas are weighted by the inverse of covariance matrix, so large $(\hat{\alpha}_i/\hat{\sigma}_{\alpha,i})^2$ dominate the *F*-statistic. This creates too much power for the *GRS* test to reject any model for large cross sections | Different magnitude of pricing errors is treated equally |
| Explanatory Power (Estimation Precision) | $RMSE(\tilde{\sigma}_\alpha) = \sqrt{\sum_{i=1}^n \tilde{\sigma}_{\alpha,i}^2/n}$ regards large standard errors highly undesirable, and penalizes models of low estimation precision | Models with large sampling errors tend to produce smaller *F*-statistics so rejected less often, i.e., the low power problem | Not considered |
| Characteristics of Good Models | Low dispersion of alphas; high estimation precision; less extreme pricing and standard errors | Low dispersion of alphas; large sampling errors | Low magnitude of alphas |

Table 2. Summary Statistics and Correlation Coefficients of Factors Returns: 1967:01 – 2016:12, 600 months

Panel A reports the mean, standard deviation, and $t$-statistic of monthly returns on ten factors, including five Fama-French factors (*MKT SMB HML RMW CMA*) of Fama and French (2015), one momentum factor (*UMD*) of Carhart (1997), three $q$ factors (*ME IA ROE*) of Hou, Xue, and Zhang (2015), and the monthly updated value factor (*HML$^m$*) of Asness and Frazzini (2013). Panel B reports the correlation coefficients of the ten factors. All data are from the French data library.

| | MKT | SMB | HML | RMW | CMA | UMD | ME | IA | ROE | HML$^m$ |
|---|---|---|---|---|---|---|---|---|---|---|
| *Panel A: Summary Statistics of Factor Returns* | | | | | | | | | | |
| Mean | 0.52 | 0.25 | 0.37 | 0.26 | 0.33 | 0.64 | 0.31 | 0.41 | 0.55 | 0.34 |
| Std | 4.53 | 3.08 | 2.88 | 2.28 | 2.03 | 4.32 | 3.08 | 1.88 | 2.55 | 3.51 |
| $t$-stat | 2.82 | 2.01 | 3.15 | 2.77 | 3.99 | 3.65 | 2.47 | 5.35 | 5.25 | 2.37 |
| *Panel B: Correlation Coefficients of Factor Returns* | | | | | | | | | | |
| MKT | 1.00 | | | | | | | | | |
| SMB | 0.28 | 1.00 | | | | | | | | |
| HML | -0.27 | -0.08 | 1.00 | | | | | | | |
| RMW | -0.24 | -0.37 | 0.09 | 1.00 | | | | | | |
| CMA | -0.40 | -0.09 | 0.70 | -0.02 | 1.00 | | | | | |
| UMD | -0.14 | -0.05 | -0.19 | 0.11 | 0.00 | 1.00 | | | | |
| ME | 0.27 | 0.97 | -0.04 | -0.37 | -0.05 | -0.02 | 1.00 | | | |
| IA | -0.38 | -0.19 | 0.67 | 0.10 | 0.91 | 0.03 | -0.15 | 1.00 | | |
| ROE | -0.20 | -0.37 | -0.14 | 0.67 | -0.09 | 0.50 | -0.31 | 0.04 | 1.00 | |
| HML$^m$ | -0.12 | -0.01 | 0.78 | -0.05 | 0.51 | -0.65 | 0.00 | 0.49 | -0.45 | 1.00 |

Table 3. Performance Metrics for 5×5 Sorted Portfolios: 1967:01 – 2016:12, 600 Observations

This table reports the distance-based metrics, the *GRS* statistic, and alpha-based statistics generated by various models for four sets of 5×5 sorted portfolios in Panels A – D, respectively. $\tilde{\alpha}_i$ and $\tilde{\sigma}_{\alpha,i}$ are the posterior estimates of pricing error and standard error for asset *i* with non-informative priors. $TD = \sqrt{\sum_{i=1}^{n}(\tilde{\alpha}_i^2 + \tilde{\sigma}_{\alpha,i}^2)}$ is the total distance between the pure mode-implied posterior distribution ($\sigma_\alpha = 0$) and the pure data-based posterior distribution ($\sigma_\alpha = \infty$). $AD = \sqrt{\sum_{i=1}^{n}(\tilde{\alpha}_i^2 + \tilde{\sigma}_{\alpha,i}^2)/n}$ is the average distance. $RMSE(\tilde{\alpha}) = \sqrt{\sum_{i=1}^{n}\tilde{\alpha}_i^2/n}$ is the square root of the mean square pricing error. $RMSE(\tilde{\sigma}_\alpha) = \sqrt{\sum_{i=1}^{n}\tilde{\sigma}_{\alpha,i}^2/n}$ is the square root of the mean square standard error. $A\tilde{\sigma}_\alpha^2/A\tilde{\alpha}^2 = MSE(\tilde{\sigma}_\alpha)/MSE(\tilde{\alpha})$ is the ratio of the mean square standard error to the mean square pricing error $A|\tilde{\alpha}| = \sum_{i=1}^{n}|\tilde{\alpha}_i|/n$ is the mean absolute pricing error (*MAE*). $A|\tilde{\alpha}|/A|\tilde{r}|$ measures the proportion of unexplained returns. *GRS* is the *F*-statistic from the finite sample *GRS* test. $R^2$ is the cross-sectional average of *R*-square values from time-series regressions. ***, **, * denote significance at 1%, 5%, and 10%, respectively.

| Model (Factors) | TD | AD | RMSE($\tilde{\alpha}$) | RMSE($\tilde{\sigma}_\alpha$) | $A\tilde{\sigma}_\alpha^2/A\tilde{\alpha}^2$ | GRS | $A|\tilde{\alpha}|$ | $A|\tilde{\alpha}|/A|\tilde{r}|$ | $R^2$ |
|---|---|---|---|---|---|---|---|---|---|
| *Panel A: 5*5 Size-B/M* | | | | | | | | | |
| CAPM (MKT) | 1.633 | 0.327 | 0.303 | 0.122 | 16% | 4.72*** | 0.255 | 144% | 75% |
| FF3 (MKT SMB HML) | 0.822 | 0.164 | 0.150 | 0.068 | 21% | 3.97*** | 0.108 | 61% | 91% |
| FF4 (MKT SMB HML UMD) | 0.751 | 0.150 | 0.134 | 0.069 | 27% | 3.38*** | 0.097 | 55% | 91% |
| FF5 (MKT SMB HML RMW CMA) | 0.727 | 0.145 | 0.129 | 0.067 | 27% | 3.30*** | 0.098 | 55% | 92% |
| FF6 (MKT SMB HML RMW CMA UMD) | 0.665 | 0.133 | 0.114 | 0.068 | 35% | 2.93*** | 0.091 | 51% | 92% |
| FF6-HML (MKT SMB RMW CMA UMD) | 0.705 | 0.141 | 0.118 | 0.078 | 43% | 2.95*** | 0.094 | 53% | 90% |
| q-factor (MKT ME IA ROE) | 0.801 | 0.160 | 0.136 | 0.085 | 39% | 3.36*** | 0.111 | 63% | 88% |
| BS (MKT SMB HML$^m$ IA ROE UMD) | 1.061 | 0.212 | 0.198 | 0.075 | 14% | 4.75*** | 0.169 | 95% | 91% |
| *Panel B: 5*5 Size-OP* | | | | | | | | | |
| CAPM (MKT) | 1.265 | 0.253 | 0.227 | 0.111 | 24% | 2.51*** | 0.202 | 129% | 79% |
| FF3 (MKT SMB HML) | 0.797 | 0.159 | 0.142 | 0.071 | 25% | 2.48*** | 0.115 | 73% | 91% |
| FF4 (MKT SMB HML UMD) | 0.745 | 0.149 | 0.130 | 0.073 | 31% | 2.22*** | 0.106 | 68% | 91% |
| FF5 (MKT SMB HML RMW CMA) | 0.491 | 0.098 | 0.075 | 0.063 | 69% | 1.98*** | 0.062 | 40% | 93% |
| FF6 (MKT SMB HML RMW CMA UMD) | 0.488 | 0.098 | 0.074 | 0.063 | 73% | 1.89*** | 0.061 | 39% | 93% |
| FF6-HML (MKT SMB RMW CMA UMD) | 0.493 | 0.099 | 0.074 | 0.065 | 76% | 1.90*** | 0.060 | 38% | 93% |
| q-factor (MKT ME IA ROE) | 0.521 | 0.104 | 0.070 | 0.077 | 119% | 1.48* | 0.058 | 37% | 91% |
| BS (MKT SMB HML$^m$ IA ROE UMD) | 0.733 | 0.147 | 0.128 | 0.072 | 31% | 2.28*** | 0.112 | 71% | 92% |
| *Panel C: 5*5 Size-INV* | | | | | | | | | |
| CAPM (MKT) | 1.532 | 0.306 | 0.285 | 0.112 | 15% | 5.11*** | 0.263 | 158% | 79% |
| FF3 (MKT SMB HML) | 0.839 | 0.168 | 0.155 | 0.065 | 17% | 4.34*** | 0.117 | 70% | 92% |
| FF4 (MKT SMB HML UMD) | 0.783 | 0.157 | 0.142 | 0.066 | 21% | 3.69*** | 0.110 | 66% | 92% |
| FF5 (MKT SMB HML RMW CMA) | 0.631 | 0.126 | 0.110 | 0.061 | 31% | 3.18*** | 0.081 | 49% | 93% |
| FF6 (MKT SMB HML RMW CMA UMD) | 0.596 | 0.119 | 0.102 | 0.062 | 37% | 2.94*** | 0.074 | 44% | 93% |
| FF6-HML (MKT SMB RMW CMA UMD) | 0.613 | 0.123 | 0.105 | 0.063 | 36% | 2.98*** | 0.078 | 47% | 93% |
| q-factor (MKT ME IA ROE) | 0.675 | 0.135 | 0.116 | 0.069 | 36% | 3.14*** | 0.084 | 50% | 92% |
| BS (MKT SMB HML$^m$ IA ROE UMD) | 0.696 | 0.139 | 0.123 | 0.066 | 29% | 3.08*** | 0.091 | 55% | 93% |
| *Panel D: 5*5 Size-MOM* | | | | | | | | | |
| CAPM (MKT) | 2.056 | 0.411 | 0.389 | 0.132 | 11% | 4.96*** | 0.326 | 121% | 74% |
| FF3 (MKT SMB HML) | 2.152 | 0.430 | 0.418 | 0.102 | 6% | 4.82*** | 0.321 | 120% | 85% |
| FF4 (MKT SMB HML UMD) | 0.892 | 0.178 | 0.163 | 0.073 | 20% | 3.69*** | 0.131 | 49% | 91% |
| FF5 (MKT SMB HML RMW CMA) | 1.744 | 0.349 | 0.333 | 0.102 | 9% | 3.99*** | 0.269 | 100% | 86% |
| FF6 (MKT SMB HML RMW CMA UMD) | 0.811 | 0.162 | 0.145 | 0.073 | 25% | 3.34*** | 0.113 | 42% | 92% |
| FF6-HML (MKT SMB RMW CMA UMD) | 0.823 | 0.165 | 0.147 | 0.074 | 25% | 3.37*** | 0.117 | 43% | 92% |
| q-factor (MKT ME IA ROE) | 0.948 | 0.190 | 0.158 | 0.104 | 43% | 2.77*** | 0.113 | 42% | 85% |
| BS (MKT SMB HML$^m$ IA ROE UMD) | 0.882 | 0.176 | 0.159 | 0.076 | 23% | 3.31*** | 0.135 | 50% | 92% |

Table 4. Pricing Errors, Standard Errors, *t*-statistics, and Marginal Distance for 5×5 Sorted Portfolios: 1967:01 – 2016:12

This table reports the posterior estimates of alphas ($\tilde{\alpha}$), standard errors ($\tilde{\sigma}_\alpha$), *t*-statistics ($\tilde{\alpha}/\tilde{\sigma}_\alpha$), and marginal distance ($d_i = \sqrt{\tilde{\alpha}_i^2 + \sigma_{\alpha,i}^2}$) with non-informative priors for each portfolio in the 5×5 *Size-B/M* sorts (Panel A), 5×5 *Size-OP* sorts (Panel B), and 5×5 *Size-MOM* sorts (Panel C). Panel A compares the *FF3* and *q*-factor models. Panel B compares the *FF5* and *q*-factor models. Panel C compares the *FF6* and *q*-factor models. $AD = \sqrt{\sum_{i=1}^n (\tilde{\alpha}_i^2 + \tilde{\sigma}_{\alpha,i}^2)/n}$ is the average distance. $RMSE(\tilde{\alpha}) = \sqrt{\sum_{i=1}^n \tilde{\alpha}_i^2/n}$ is the square root of the mean square pricing error. $RMSE(\tilde{\sigma}_\alpha) = \sqrt{\sum_{i=1}^n \tilde{\sigma}_{\alpha,i}^2/n}$ is the square root of the mean square standard error. $A|\tilde{\alpha}| = \sum_{i=1}^n |\tilde{\alpha}_i|/n$ is the mean absolute pricing error (*MAE*). *GRS* is the *F*-statistic from the finite sample *GRS* test.

| | | | | | | | | | | | *Panel A: 25 Size-B/M portfolios* | | | | | | | | | |
|---|---|---|---|---|---|---|---|---|---|---|---|---|---|---|---|---|---|---|---|---|
| **FF3** | | $\tilde{\alpha}$ | | | | | $\tilde{\sigma}_\alpha$ | | | | | $t_i$ | | | | | $d_i$ | | | |
| B/M→ | Low | 2 | 3 | 4 | High | Low | 2 | 3 | 4 | High | Low | 2 | 3 | 4 | High | Low | 2 | 3 | 4 | High |
| Small | -0.52 | 0.02 | -0.01 | 0.20 | 0.14 | 0.10 | 0.07 | 0.05 | 0.06 | 0.06 | -5.27 | 0.22 | -0.20 | 3.58 | 2.30 | 0.53 | 0.07 | 0.05 | 0.21 | 0.15 |
| 2 | -0.17 | 0.02 | 0.08 | 0.12 | -0.03 | 0.06 | 0.05 | 0.06 | 0.05 | 0.06 | -2.69 | 0.28 | 1.46 | 2.26 | -0.57 | 0.18 | 0.06 | 0.10 | 0.13 | 0.06 |
| 3 | -0.06 | 0.08 | -0.02 | 0.07 | 0.10 | 0.06 | 0.07 | 0.07 | 0.06 | 0.08 | -1.05 | 1.24 | -0.32 | 1.14 | 1.25 | 0.09 | 0.10 | 0.07 | 0.10 | 0.12 |
| 4 | 0.11 | -0.08 | -0.04 | 0.08 | -0.12 | 0.06 | 0.07 | 0.07 | 0.07 | 0.09 | 1.75 | -1.13 | -0.57 | 1.15 | -1.45 | 0.13 | 0.11 | 0.08 | 0.10 | 0.15 |
| Big | 0.16 | 0.04 | -0.02 | -0.24 | -0.16 | 0.05 | 0.06 | 0.07 | 0.07 | 0.10 | 3.38 | 0.70 | -0.21 | -3.64 | -1.55 | 0.17 | 0.07 | 0.07 | 0.25 | 0.19 |
| | *RMSE* ($\tilde{\alpha}$)=0.150; *RMSE* ($\tilde{\sigma}_\alpha$)=0.068; *AD*=0.164; *A* |$\tilde{\alpha}$|=0.108; *GRS*=3.97 | | | | | | | | | | | | | | | | | | | |
| **q-factor** | | $\tilde{\alpha}$ | | | | | $\tilde{\sigma}_\alpha$ | | | | | $t_i$ | | | | | $d_i$ | | | |
| OP→ | Low | 2 | 3 | 4 | High | Low | 2 | 3 | 4 | High | Low | 2 | 3 | 4 | High | Low | 2 | 3 | 4 | High |
| Small | -0.23 | 0.23 | 0.06 | 0.26 | 0.25 | 0.12 | 0.08 | 0.06 | 0.07 | 0.08 | -1.97 | 3.03 | 0.97 | 3.79 | 3.02 | 0.26 | 0.24 | 0.09 | 0.27 | 0.26 |
| 2 | -0.08 | 0.03 | 0.01 | 0.10 | 0.03 | 0.09 | 0.06 | 0.07 | 0.07 | 0.09 | -0.96 | 0.43 | 0.14 | 1.32 | 0.30 | 0.12 | 0.07 | 0.07 | 0.12 | 0.09 |
| 3 | 0.02 | -0.01 | -0.08 | 0.05 | 0.12 | 0.08 | 0.07 | 0.08 | 0.08 | 0.11 | 0.23 | -0.13 | -0.99 | 0.58 | 1.16 | 0.08 | 0.07 | 0.11 | 0.10 | 0.16 |
| 4 | 0.14 | -0.20 | -0.12 | 0.08 | -0.02 | 0.08 | 0.08 | 0.08 | 0.08 | 0.12 | 1.84 | -2.67 | -1.41 | 0.93 | -0.19 | 0.16 | 0.22 | 0.15 | 0.11 | 0.12 |
| Big | 0.09 | -0.09 | -0.11 | -0.22 | 0.15 | 0.06 | 0.06 | 0.08 | 0.09 | 0.14 | 1.54 | -1.56 | -1.35 | -2.34 | 1.08 | 0.10 | 0.11 | 0.14 | 0.24 | 0.20 |
| | *RMSE* ($\tilde{\alpha}$)=0.136; *RMSE* ($\tilde{\sigma}_\alpha$)=0.085; *AD*=0.160; *A* |$\tilde{\alpha}$|=0.111; *GRS*=3.36 | | | | | | | | | | | | | | | | | | | |

Table 4 (cont.)

### Panel B: 25 Size-OP portfolios

**FF5**

| OP→ | $\tilde{\alpha}$ | | | | | $\tilde{\sigma}_\alpha$ | | | | | $t_i$ | | | | | $d_i$ | | | | |
|---|---|---|---|---|---|---|---|---|---|---|---|---|---|---|---|---|---|---|---|---|
| | Low | 2 | 3 | 4 | High | Low | 2 | 3 | 4 | High | Low | 2 | 3 | 4 | High | Low | 2 | 3 | 4 | High |
| Small | -0.13 | 0.08 | -0.06 | -0.02 | -0.15 | 0.08 | 0.06 | 0.06 | 0.06 | 0.07 | -1.69 | 1.32 | -1.02 | -0.33 | -2.15 | 0.15 | 0.10 | 0.09 | 0.07 | 0.17 |
| 2 | -0.07 | -0.10 | -0.01 | -0.01 | 0.00 | 0.06 | 0.06 | 0.05 | 0.05 | 0.06 | -1.09 | -1.70 | -0.22 | -0.24 | 0.06 | 0.09 | 0.11 | 0.05 | 0.05 | 0.06 |
| 3 | 0.05 | -0.02 | -0.03 | -0.08 | 0.02 | 0.07 | 0.07 | 0.06 | 0.06 | 0.07 | 0.65 | -0.27 | -0.63 | -1.45 | 0.31 | 0.09 | 0.07 | 0.07 | 0.10 | 0.07 |
| 4 | 0.11 | 0.07 | -0.15 | -0.09 | 0.01 | 0.08 | 0.07 | 0.06 | 0.06 | 0.07 | 1.34 | 0.97 | -2.39 | -1.49 | 0.14 | 0.14 | 0.09 | 0.17 | 0.11 | 0.07 |
| Big | 0.09 | -0.06 | -0.03 | 0.05 | 0.05 | 0.08 | 0.06 | 0.06 | 0.05 | 0.04 | 1.24 | -1.11 | -0.44 | 1.06 | 1.25 | 0.12 | 0.08 | 0.06 | 0.07 | 0.07 |

RMSE ($\tilde{\alpha}$)=0.075; RMSE ($\tilde{\sigma}_\alpha$)=0.063; AD=0.098; A |$\tilde{\alpha}$|=0.062; GRS=1.98

**q-factor**

| OP→ | $\tilde{\alpha}$ | | | | | $\tilde{\sigma}_\alpha$ | | | | | $t_i$ | | | | | $d_i$ | | | | |
|---|---|---|---|---|---|---|---|---|---|---|---|---|---|---|---|---|---|---|---|---|
| | Low | 2 | 3 | 4 | High | Low | 2 | 3 | 4 | High | Low | 2 | 3 | 4 | High | Low | 2 | 3 | 4 | High |
| Small | 0.01 | 0.14 | -0.02 | 0.04 | -0.06 | 0.09 | 0.08 | 0.08 | 0.09 | 0.09 | 0.13 | 1.80 | -0.29 | 0.45 | -0.65 | 0.09 | 0.16 | 0.08 | 0.09 | 0.11 |
| 2 | -0.03 | -0.06 | 0.00 | 0.04 | 0.03 | 0.08 | 0.07 | 0.07 | 0.07 | 0.09 | -0.39 | -0.91 | 0.06 | 0.56 | 0.40 | 0.09 | 0.09 | 0.07 | 0.08 | 0.09 |
| 3 | 0.12 | -0.03 | -0.03 | -0.05 | 0.03 | 0.09 | 0.07 | 0.06 | 0.07 | 0.09 | 1.31 | -0.42 | -0.47 | -0.75 | 0.30 | 0.15 | 0.08 | 0.07 | 0.09 | 0.09 |
| 4 | 0.16 | 0.08 | -0.12 | -0.08 | 0.01 | 0.10 | 0.07 | 0.07 | 0.07 | 0.07 | 1.61 | 1.12 | -1.65 | -1.09 | 0.10 | 0.19 | 0.11 | 0.14 | 0.10 | 0.07 |
| Big | 0.04 | -0.06 | -0.08 | 0.07 | 0.05 | 0.10 | 0.06 | 0.06 | 0.05 | 0.05 | 0.38 | -0.94 | -1.28 | 1.48 | 1.03 | 0.11 | 0.09 | 0.10 | 0.09 | 0.07 |

RMSE ($\tilde{\alpha}$)=0.070; RMSE ($\tilde{\sigma}_\alpha$)=0.077; AD=0.104; A |$\tilde{\alpha}$|=0.058; GRS=1.48

### Panel C: 25 Size-MOM portfolios

**FF6**

| UMD→ | $\tilde{\alpha}$ | | | | | $\tilde{\sigma}_\alpha$ | | | | | $t_i$ | | | | | $d_i$ | | | | |
|---|---|---|---|---|---|---|---|---|---|---|---|---|---|---|---|---|---|---|---|---|
| | Low | 2 | 3 | 4 | High | Low | 2 | 3 | 4 | High | Low | 2 | 3 | 4 | High | Low | 2 | 3 | 4 | High |
| Small | -0.23 | -0.03 | 0.11 | 0.19 | 0.40 | 0.10 | 0.06 | 0.06 | 0.07 | 0.08 | -2.32 | -0.43 | 1.67 | 2.88 | 4.78 | 0.25 | 0.07 | 0.12 | 0.20 | 0.41 |
| 2 | -0.10 | 0.04 | 0.03 | 0.12 | 0.22 | 0.08 | 0.06 | 0.06 | 0.06 | 0.07 | -1.33 | 0.66 | 0.49 | 2.23 | 3.44 | 0.13 | 0.07 | 0.06 | 0.13 | 0.23 |
| 3 | 0.12 | 0.03 | 0.00 | -0.12 | 0.19 | 0.09 | 0.06 | 0.06 | 0.06 | 0.07 | 1.32 | 0.44 | 0.01 | -1.87 | 2.90 | 0.15 | 0.07 | 0.06 | 0.13 | 0.20 |
| 4 | 0.12 | 0.05 | 0.01 | -0.03 | 0.05 | 0.10 | 0.07 | 0.06 | 0.06 | 0.07 | 1.13 | 0.68 | 0.10 | -0.49 | 0.66 | 0.15 | 0.08 | 0.06 | 0.07 | 0.09 |
| Big | 0.14 | 0.21 | -0.10 | -0.15 | -0.04 | 0.11 | 0.07 | 0.06 | 0.06 | 0.07 | 1.30 | 3.03 | -1.69 | -2.58 | -0.65 | 0.18 | 0.22 | 0.12 | 0.17 | 0.08 |

RMSE ($\tilde{\alpha}$)=0.145; RMSE ($\tilde{\sigma}_\alpha$)=0.073; AD=0.162; A |$\tilde{\alpha}$|=0.113; GRS=3.34

**q-factor**

| UMD→ | $\tilde{\alpha}$ | | | | | $\tilde{\sigma}_\alpha$ | | | | | $t_i$ | | | | | $d_i$ | | | | |
|---|---|---|---|---|---|---|---|---|---|---|---|---|---|---|---|---|---|---|---|---|
| | Low | 2 | 3 | 4 | High | Low | 2 | 3 | 4 | High | Low | 2 | 3 | 4 | High | Low | 2 | 3 | 4 | High |
| Small | -0.15 | 0.01 | 0.16 | 0.27 | 0.50 | 0.14 | 0.09 | 0.08 | 0.08 | 0.11 | -1.09 | 0.13 | 2.00 | 3.48 | 4.68 | 0.21 | 0.09 | 0.18 | 0.28 | 0.51 |
| 2 | -0.11 | 0.02 | 0.03 | 0.15 | 0.28 | 0.13 | 0.09 | 0.07 | 0.07 | 0.09 | -0.79 | 0.17 | 0.38 | 2.25 | 2.99 | 0.17 | 0.09 | 0.08 | 0.16 | 0.29 |
| 3 | 0.07 | -0.02 | -0.04 | -0.11 | 0.23 | 0.15 | 0.10 | 0.08 | 0.07 | 0.10 | 0.43 | -0.17 | -0.55 | -1.57 | 2.42 | 0.17 | 0.10 | 0.09 | 0.13 | 0.25 |
| 4 | 0.04 | -0.06 | -0.03 | -0.02 | 0.10 | 0.17 | 0.11 | 0.08 | 0.07 | 0.10 | 0.24 | -0.54 | -0.31 | -0.33 | 0.92 | 0.17 | 0.12 | 0.09 | 0.07 | 0.14 |
| Big | 0.03 | 0.08 | -0.14 | -0.15 | 0.03 | 0.17 | 0.11 | 0.07 | 0.07 | 0.10 | 0.15 | 0.75 | -1.99 | -2.27 | 0.25 | 0.17 | 0.14 | 0.16 | 0.16 | 0.11 |

RMSE ($\tilde{\alpha}$)=0.158; RMSE ($\tilde{\sigma}_\alpha$)=0.104; AD=0.190; A |$\tilde{\alpha}$|=0.113; GRS=2.77

Table 5. Performance Metrics for 2×4×4 Sorted Portfolios: 1967:01 – 2016:12, 600 Observations

This table reports the distance-based metrics, the *GRS* statistic, and alpha-based statistics generated by various models for three sets of 2×4×4 sorted portfolios in Panels A – C, respectively. $\tilde{\alpha}_i$ and $\tilde{\sigma}_{\alpha,i}$ are the posterior estimates of pricing error and standard error for asset $i$ with non-informative priors. $TD = \sqrt{\sum_{i=1}^{n}(\tilde{\alpha}_i^2 + \tilde{\sigma}_{\alpha,i}^2)}$ is the total distance between the pure mode-implied posterior distribution ($\sigma_\alpha = 0$) and the pure data-based posterior distribution ($\sigma_\alpha = \infty$). $AD = \sqrt{\sum_{i=1}^{n}(\tilde{\alpha}_i^2 + \tilde{\sigma}_{\alpha,i}^2)/n}$ is the average distance. $RMSE(\tilde{\alpha}) = \sqrt{\sum_{i=1}^{n}\tilde{\alpha}_i^2/n}$ is the square root of the mean square pricing error. $RMSE(\tilde{\sigma}_\alpha) = \sqrt{\sum_{i=1}^{n}\tilde{\sigma}_{\alpha,i}^2/n}$ is the square root of the mean square standard error. $A\tilde{\sigma}_\alpha^2/A\tilde{\alpha}^2 = MSE(\tilde{\sigma}_\alpha)/MSE(\tilde{\alpha})$ is the ratio of the mean square standard errors to the mean square pricing error. $A|\tilde{\alpha}| = \sum_{i=1}^{n}|\tilde{\alpha}_i|/n$ is the mean absolute pricing error (*MAE*). $A|\tilde{\alpha}|/A|\tilde{r}|$ measures the proportion of unexplained returns. *GRS* is the *F*-statistic from the finite sample *GRS* test. $R^2$ is the cross-sectional average of *R*-square values from time-series regressions. ***, **, * denote significance at 1%, 5%, and 10%, respectively.

| | Panel A: 2*4*4 Size-B/M-OP | | | | | | | | |
|---|---|---|---|---|---|---|---|---|---|
| Model (Factors) | TD | AD | RMSE($\tilde{\alpha}$) | RMSE($\tilde{\sigma}_\alpha$) | $A\tilde{\sigma}_\alpha^2/A\tilde{\alpha}^2$ | GRS | $A|\tilde{\alpha}|$ | $A|\tilde{\alpha}|/A|\tilde{r}|$ | $R^2$ |
| CAPM (MKT) | 2.199 | 0.389 | 0.363 | 0.139 | 15% | 2.80*** | 0.286 | 124% | 70% |
| FF3 (MKT SMB HML) | 1.216 | 0.215 | 0.189 | 0.102 | 29% | 2.47*** | 0.141 | 61% | 84% |
| FF4 (MKT SMB HML UMD) | 1.205 | 0.213 | 0.186 | 0.104 | 31% | 2.10*** | 0.138 | 60% | 84% |
| FF5 (MKT SMB HML RMW CMA) | 0.990 | 0.175 | 0.144 | 0.099 | 47% | 2.18*** | 0.111 | 48% | 86% |
| FF6 (MKT SMB HML RMW CMA UMD) | 0.979 | 0.173 | 0.141 | 0.100 | 50% | 2.02*** | 0.111 | 48% | 86% |
| FF6-HML (MKT SMB RMW CMA UMD) | 1.020 | 0.180 | 0.145 | 0.108 | 55% | 1.92*** | 0.111 | 48% | 84% |
| q-factor (MKT ME IA ROE) | 1.139 | 0.201 | 0.164 | 0.117 | 51% | 1.89*** | 0.127 | 55% | 81% |
| BS (MKT SMB HML$^m$ IA ROE UMD) | 1.412 | 0.250 | 0.224 | 0.109 | 24% | 3.26*** | 0.185 | 80% | 85% |
| | Panel B: 2*4*4 Size-B/M-INV | | | | | | | | |
| Model (Factors) | TD | AD | RMSE($\tilde{\alpha}$) | RMSE($\tilde{\sigma}_\alpha$) | $A\tilde{\sigma}_\alpha^2/A\tilde{\alpha}^2$ | GRS | $A|\tilde{\alpha}|$ | $A|\tilde{\alpha}|/A|\tilde{r}|$ | $R^2$ |
| CAPM (MKT) | 1.834 | 0.324 | 0.300 | 0.122 | 16% | 3.94*** | 0.251 | 134% | 73% |
| FF3 (MKT SMB HML) | 1.046 | 0.185 | 0.167 | 0.079 | 22% | 3.47*** | 0.144 | 77% | 87% |
| FF4 (MKT SMB HML UMD) | 0.971 | 0.172 | 0.152 | 0.081 | 28% | 2.86*** | 0.130 | 69% | 87% |
| FF5 (MKT SMB HML RMW CMA) | 0.864 | 0.153 | 0.132 | 0.077 | 34% | 2.52*** | 0.106 | 57% | 89% |
| FF6 (MKT SMB HML RMW CMA UMD) | 0.814 | 0.144 | 0.121 | 0.078 | 41% | 2.19*** | 0.098 | 53% | 89% |
| FF6-HML (MKT SMB RMW CMA UMD) | 0.818 | 0.145 | 0.116 | 0.086 | 55% | 2.14*** | 0.094 | 50% | 86% |
| q-factor (MKT ME IA ROE) | 0.921 | 0.163 | 0.134 | 0.092 | 47% | 2.36*** | 0.105 | 56% | 85% |
| BS (MKT SMB HML$^m$ IA ROE UMD) | 1.325 | 0.234 | 0.218 | 0.085 | 15% | 3.71*** | 0.179 | 96% | 88% |
| | Panel C: 2*4*4 Size-OP-INV | | | | | | | | |
| Model (Factors) | TD | AD | RMSE($\tilde{\alpha}$) | RMSE($\tilde{\sigma}_\alpha$) | $A\tilde{\sigma}_\alpha^2/A\tilde{\alpha}^2$ | GRS | $A|\tilde{\alpha}|$ | $A|\tilde{\alpha}|/A|\tilde{r}|$ | $R^2$ |
| CAPM (MKT) | 2.055 | 0.363 | 0.344 | 0.116 | 11% | 5.25*** | 0.274 | 124% | 75% |
| FF3 (MKT SMB HML) | 1.458 | 0.258 | 0.244 | 0.082 | 11% | 4.72*** | 0.188 | 85% | 87% |
| FF4 (MKT SMB HML UMD) | 1.335 | 0.236 | 0.221 | 0.083 | 14% | 3.94*** | 0.169 | 77% | 87% |
| FF5 (MKT SMB HML RMW CMA) | 0.975 | 0.172 | 0.155 | 0.075 | 23% | 3.63*** | 0.116 | 53% | 89% |
| FF6 (MKT SMB HML RMW CMA UMD) | 0.913 | 0.161 | 0.142 | 0.076 | 28% | 3.27*** | 0.110 | 50% | 89% |
| FF6-HML (MKT SMB RMW CMA UMD) | 0.935 | 0.165 | 0.146 | 0.077 | 28% | 3.28*** | 0.115 | 52% | 89% |
| q-factor (MKT ME IA ROE) | 0.950 | 0.168 | 0.144 | 0.086 | 36% | 2.60*** | 0.113 | 51% | 87% |
| BS (MKT SMB HML$^m$ IA ROE UMD) | 1.032 | 0.182 | 0.162 | 0.083 | 26% | 3.01*** | 0.134 | 61% | 88% |

Table 6. Performance Metrics for Four Groups of Decile Portfolios: 1967:01 – 2016:12, 600 Observations

This table reports the distance-based metrics, the *GRS* statistic, and alpha-based statistics generated by various models for four groups of 15 univariate-sorted decile portfolios in Panels A – D, respectively. $\tilde{\alpha}_i$ and $\tilde{\sigma}_{\alpha,i}$ are the posterior estimates of pricing error and standard error for asset $i$ with non-informative priors. $TD = \sqrt{\sum_{i=1}^{n}(\tilde{\alpha}_i^2 + \tilde{\sigma}_{\alpha,i}^2)}$ is the total distance between the pure mode-implied posterior distribution ($\sigma_\alpha = 0$) and the pure data-based posterior distribution ($\sigma_\alpha = \infty$). $AD = \sqrt{\sum_{i=1}^{n}(\tilde{\alpha}_i^2 + \tilde{\sigma}_{\alpha,i}^2)/n}$ is the average distance. $RMSE(\tilde{\alpha}) = \sqrt{\sum_{i=1}^{n}\tilde{\alpha}_i^2/n}$ is the square root of the mean square pricing error. $RMSE(\tilde{\sigma}_\alpha) = \sqrt{\sum_{i=1}^{n}\tilde{\sigma}_{\alpha,i}^2/n}$ is the square root of the mean square standard error. $A\tilde{\sigma}_\alpha^2/A\tilde{\alpha}^2 = MSE(\tilde{\sigma}_\alpha)/MSE(\tilde{\alpha})$ is the ratio of the mean square standard error to the mean square pricing error. $A|\tilde{\alpha}| = \sum_{i=1}^{n}|\tilde{\alpha}_i|/n$ is the mean absolute pricing error (*MAE*). $A|\tilde{\alpha}|/A|\tilde{r}|$ measures the proportion of unexplained returns. *GRS* is the *F*-statistic from the finite sample *GRS* test. $R^2$ is the cross-sectional average of *R*-square values from time-series regressions. ***, **, * denote significance at 1%, 5%, and 10%, respectively.

| Model (Factors) | TD | AD | RMSE($\tilde{\alpha}$) | RMSE($\tilde{\sigma}_\alpha$) | $A\tilde{\sigma}_\alpha^2/A\tilde{\alpha}^2$ | GRS | $A\|\tilde{\alpha}\|$ | $A\|\tilde{\alpha}\|/A\|\tilde{r}\|$ | $R^2$ |
|---|---|---|---|---|---|---|---|---|---|
| *Panel A: FF-factor (10 Size + 10 B/M + 10 OP + 10 INV)* | | | | | | | | | |
| CAPM (MKT) | 1.188 | 0.188 | 0.168 | 0.083 | 24% | 2.25*** | 0.137 | 133% | 85% |
| FF3 (MKT SMB HML) | 0.827 | 0.131 | 0.116 | 0.061 | 27% | 1.94*** | 0.085 | 83% | 92% |
| FF4 (MKT SMB HML UMD) | 0.730 | 0.115 | 0.098 | 0.062 | 40% | 1.66*** | 0.074 | 72% | 92% |
| FF5 (MKT SMB HML RMW CMA) | 0.610 | 0.097 | 0.077 | 0.058 | 55% | 1.35* | 0.061 | 59% | 93% |
| FF6 (MKT SMB HML RMW CMA UMD) | 0.574 | 0.091 | 0.070 | 0.058 | 69% | 1.26 | 0.053 | 52% | 93% |
| FF6-HML (MKT SMB RMW CMA UMD) | 0.588 | 0.093 | 0.069 | 0.062 | 80% | 1.27 | 0.054 | 52% | 92% |
| q-factor (MKT ME IA ROE) | 0.681 | 0.108 | 0.085 | 0.067 | 62% | 1.73*** | 0.066 | 65% | 91% |
| BS (MKT SMB HML$^m$ IA ROE UMD) | 0.920 | 0.145 | 0.131 | 0.064 | 24% | 2.64*** | 0.097 | 94% | 92% |
| *Panel B: Valuation (10 E/P + 10 CF/P + 10 D/P)* | | | | | | | | | |
| CAPM (MKT) | 1.171 | 0.214 | 0.195 | 0.087 | 20% | 1.22 | 0.163 | 170% | 80% |
| FF3 (MKT SMB HML) | 0.535 | 0.098 | 0.062 | 0.075 | 147% | 0.99 | 0.051 | 53% | 85% |
| FF4 (MKT SMB HML UMD) | 0.570 | 0.104 | 0.070 | 0.077 | 119% | 1.11 | 0.056 | 59% | 85% |
| FF5 (MKT SMB HML RMW CMA) | 0.751 | 0.137 | 0.116 | 0.074 | 41% | 1.53** | 0.098 | 102% | 87% |
| FF6 (MKT SMB HML RMW CMA UMD) | 0.718 | 0.131 | 0.108 | 0.074 | 47% | 1.50** | 0.086 | 89% | 87% |
| FF6-HML (MKT SMB RMW CMA UMD) | 0.731 | 0.133 | 0.107 | 0.080 | 56% | 1.36* | 0.092 | 96% | 85% |
| q-factor (MKT ME IA ROE) | 0.941 | 0.172 | 0.149 | 0.086 | 33% | 1.95*** | 0.128 | 134% | 83% |
| BS (MKT SMB HML$^m$ IA ROE UMD) | 1.228 | 0.224 | 0.210 | 0.079 | 14% | 2.88*** | 0.189 | 196% | 86% |
| *Panel C: Prior Return (10 MOM + 10 STR + 10 LTR)* | | | | | | | | | |
| CAPM (MKT) | 1.468 | 0.268 | 0.249 | 0.099 | 16% | 2.69*** | 0.176 | 117% | 82% |
| FF3 (MKT SMB HML) | 1.552 | 0.283 | 0.268 | 0.093 | 12% | 2.42*** | 0.154 | 102% | 84% |
| FF4 (MKT SMB HML UMD) | 0.838 | 0.153 | 0.130 | 0.081 | 39% | 1.88*** | 0.092 | 62% | 87% |
| FF5 (MKT SMB HML RMW CMA) | 1.253 | 0.229 | 0.209 | 0.093 | 20% | 2.04*** | 0.132 | 88% | 85% |
| FF6 (MKT SMB HML RMW CMA UMD) | 0.783 | 0.143 | 0.119 | 0.080 | 45% | 1.76*** | 0.102 | 68% | 88% |
| FF6-HML (MKT SMB RMW CMA UMD) | 0.781 | 0.143 | 0.118 | 0.080 | 47% | 1.69** | 0.098 | 65% | 88% |
| q-factor (MKT ME IA ROE) | 0.835 | 0.152 | 0.119 | 0.096 | 65% | 1.48* | 0.097 | 65% | 85% |
| BS (MKT SMB HML$^m$ IA ROE UMD) | 0.880 | 0.161 | 0.137 | 0.084 | 37% | 2.29*** | 0.122 | 82% | 88% |
| *Panel D: Other Anomaly (10 AC + 10 NI + 10 Beta + 10 VAR + 10 RVAR)* | | | | | | | | | |
| CAPM (MKT) | 1.867 | 0.264 | 0.248 | 0.090 | 13% | 2.98*** | 0.164 | 152% | 85% |
| FF3 (MKT SMB HML) | 1.861 | 0.263 | 0.251 | 0.080 | 10% | 3.89*** | 0.154 | 143% | 87% |
| FF4 (MKT SMB HML UMD) | 1.465 | 0.207 | 0.191 | 0.080 | 17% | 3.38*** | 0.122 | 113% | 88% |
| FF5 (MKT SMB HML RMW CMA) | 1.276 | 0.180 | 0.163 | 0.077 | 22% | 3.27*** | 0.120 | 111% | 89% |
| FF6 (MKT SMB HML RMW CMA UMD) | 1.124 | 0.159 | 0.139 | 0.077 | 31% | 3.00*** | 0.102 | 95% | 89% |
| FF6-HML (MKT SMB RMW CMA UMD) | 1.113 | 0.157 | 0.137 | 0.077 | 32% | 2.89*** | 0.100 | 93% | 89% |
| q-factor (MKT ME IA ROE) | 1.185 | 0.168 | 0.146 | 0.082 | 32% | 2.89*** | 0.112 | 104% | 88% |
| BS (MKT SMB HML$^m$ IA ROE UMD) | 1.310 | 0.185 | 0.166 | 0.082 | 25% | 3.45*** | 0.135 | 126% | 88% |

Table 7. Pricing Errors, Standard Errors, *t*-statistics, and Marginal Distance for Ten Momentum Portfolios: 1967:01 – 2016:12, 600 Observations

This table reports the posterior estimates of alphas ($\tilde{\alpha}$), standard errors ($\tilde{\sigma}_\alpha$), *t*-statistics ($\tilde{\alpha}/\tilde{\sigma}_\alpha$), and marginal distance ($d_i = \sqrt{\tilde{\alpha}_i^2 + \sigma_{\alpha,i}^2}$) with non-informative priors for each portfolio in the decile portfolios sorted by momentum. Three models: *FF5*, *FF6-HML*, and *q*-factor are compared. $AD = \sqrt{\sum_{i=1}^{n}(\tilde{\alpha}_i^2 + \tilde{\sigma}_{\alpha,i}^2)/n}$ is the average distance. $RMSE(\tilde{\alpha}) = \sqrt{\sum_{i=1}^{n} \tilde{\alpha}_i^2/n}$ is the square root of the mean square pricing error. $RMSE(\tilde{\sigma}_\alpha) = \sqrt{\sum_{i=1}^{n} \tilde{\sigma}_{\alpha,i}^2/n}$ is the square root of the mean square standard error. $A|\tilde{\alpha}| = \sum_{i=1}^{n}|\tilde{\alpha}_i|/n$ is the mean absolute pricing error (*MAE*). *GRS* is the *F*-statistic from the finite sample *GRS* test.

| **FF5** | Lo 10 | 02-Dec | 03-Dec | 04-Dec | 05-Dec | 06-Dec | 07-Dec | 08-Dec | 09-Dec | Hi 10 |
|---|---|---|---|---|---|---|---|---|---|---|
| $\tilde{\alpha}$ | -0.78 | -0.33 | -0.21 | -0.14 | -0.18 | -0.17 | -0.11 | 0.03 | 0.07 | 0.56 |
| $\tilde{\sigma}_\alpha$ | 0.20 | 0.14 | 0.11 | 0.09 | 0.07 | 0.07 | 0.06 | 0.07 | 0.08 | 0.12 |
| $t_i$ | -3.92 | -2.40 | -1.83 | -1.63 | -2.53 | -2.53 | -1.75 | 0.47 | 0.81 | 4.48 |
| $d_i$ | 0.81 | 0.36 | 0.24 | 0.17 | 0.19 | 0.18 | 0.13 | 0.08 | 0.11 | 0.57 |
| *RMSE* ($\tilde{\alpha}$)=0.343; *RMSE* ($\tilde{\sigma}_\alpha$)=0.110; *AD*=0.360; *A*|$\tilde{\alpha}$|=0.258; *GRS*=3.95 | | | | | | | | | | |
| **FF6-HML** | Lo 10 | 02-Dec | 03-Dec | 04-Dec | 05-Dec | 06-Dec | 07-Dec | 08-Dec | 09-Dec | Hi 10 |
| $\tilde{\alpha}$ | -0.12 | 0.16 | 0.18 | 0.09 | -0.04 | -0.12 | -0.15 | -0.11 | -0.15 | 0.16 |
| $\tilde{\sigma}_\alpha$ | 0.11 | 0.07 | 0.06 | 0.07 | 0.06 | 0.07 | 0.06 | 0.06 | 0.06 | 0.08 |
| $t_i$ | -1.05 | 2.34 | 2.71 | 1.42 | -0.65 | -1.76 | -2.33 | -1.87 | -2.33 | 2.09 |
| $d_i$ | 0.17 | 0.17 | 0.19 | 0.11 | 0.08 | 0.13 | 0.16 | 0.12 | 0.16 | 0.18 |
| *RMSE* ($\tilde{\alpha}$)=0.133; *RMSE* ($\tilde{\sigma}_\alpha$)=0.072; *AD*=0.151; *A*|$\tilde{\alpha}$|=0.127; *GRS*=2.85 | | | | | | | | | | |
| **q-factor** | Lo 10 | 02-Dec | 03-Dec | 04-Dec | 05-Dec | 06-Dec | 07-Dec | 08-Dec | 09-Dec | Hi 10 |
| $\tilde{\alpha}$ | -0.18 | 0.05 | 0.05 | -0.02 | -0.11 | -0.15 | -0.16 | -0.11 | -0.14 | 0.24 |
| $\tilde{\sigma}_\alpha$ | 0.19 | 0.14 | 0.12 | 0.09 | 0.08 | 0.07 | 0.07 | 0.07 | 0.08 | 0.13 |
| $t_i$ | -0.98 | 0.40 | 0.39 | -0.17 | -1.40 | -2.05 | -2.25 | -1.63 | -1.77 | 1.94 |
| $d_i$ | 0.26 | 0.15 | 0.13 | 0.10 | 0.13 | 0.17 | 0.17 | 0.13 | 0.16 | 0.27 |
| *RMSE* ($\tilde{\alpha}$)=0.137; *RMSE* ($\tilde{\sigma}_\alpha$)=0.109; *AD*=0.175; *A*|$\tilde{\alpha}$|=0.121; *GRS*=2.23 | | | | | | | | | | |

Table 8. Performance Metrics for Two Board Cross Sections: 1967:01 – 2016:12, 600 Observations

This table reports the distance-based metrics, the *GRS* statistic, and alpha-based statistics generated by various models for two augmented cross sections in Panels A – B, respectively. $\tilde{\alpha}_i$ and $\tilde{\sigma}_{\alpha,i}$ are the posterior estimates of pricing error and standard error for asset $i$ with non-informative priors. $TD = \sqrt{\sum_{i=1}^{n}(\tilde{\alpha}_i^2 + \tilde{\sigma}_{\alpha,i}^2)}$ is the total distance between the pure mode-implied posterior distribution ($\sigma_\alpha = 0$) and the pure data-based posterior distribution ($\sigma_\alpha = \infty$). $AD = \sqrt{\sum_{i=1}^{n}(\tilde{\alpha}_i^2 + \tilde{\sigma}_{\alpha,i}^2)/n}$ is the average distance. $RMSE(\tilde{\alpha}) = \sqrt{\sum_{i=1}^{n}\tilde{\alpha}_i^2/n}$ is the square root of the mean square pricing error. $RMSE(\tilde{\sigma}_\alpha) = \sqrt{\sum_{i=1}^{n}\tilde{\sigma}_{\alpha,i}^2/n}$ is the square root of the mean square standard error. $A\tilde{\sigma}_\alpha^2/A\tilde{\alpha}^2 = MSE(\tilde{\sigma}_\alpha)/MSE(\tilde{\alpha})$ is the ratio of the mean square standard error to the mean square pricing error. $A|\tilde{\alpha}| = \sum_{i=1}^{n}|\tilde{\alpha}_i|/n$ is the mean absolute pricing error (*MAE*). $A|\tilde{\alpha}|/A|\tilde{r}|$ measures the proportion of unexplained returns. *GRS* is the *F*-statistic from the finite sample *GRS* test. $R^2$ is the cross-sectional average of *R*-square values from time-series regressions. ***, **, * denote significance at 1%, 5%, and 10%, respectively.

| Model (Factors) | TD | AD | RMSE($\tilde{\alpha}$) | RMSE($\tilde{\sigma}_\alpha$) | $A\tilde{\sigma}_\alpha^2/A\tilde{\alpha}^2$ | GRS | $A|\tilde{\alpha}|$ | $A|\tilde{\alpha}|/A|\tilde{r}|$ | $R^2$ |
|---|---|---|---|---|---|---|---|---|---|
| **Panel A: 4\*25+3\*32** | | | | | | | | | |
| CAPM (MKT) | 4.823 | 0.344 | 0.322 | 0.123 | 15% | 2.86*** | 0.266 | 131% | 75% |
| FF3 (MKT SMB HML) | 3.368 | 0.241 | 0.226 | 0.083 | 14% | 2.73*** | 0.162 | 79% | 88% |
| FF4 (MKT SMB HML UMD) | 2.589 | 0.185 | 0.167 | 0.080 | 23% | 2.60*** | 0.128 | 63% | 89% |
| FF5 (MKT SMB HML RMW CMA) | 2.624 | 0.187 | 0.170 | 0.080 | 22% | 2.45*** | 0.119 | 59% | 89% |
| FF6 (MKT SMB HML RMW CMA UMD) | 2.036 | 0.145 | 0.124 | 0.076 | 38% | 2.39*** | 0.095 | 47% | 90% |
| FF6-HML (MKT SMB RMW CMA UMD) | 2.092 | 0.149 | 0.126 | 0.081 | 42% | 2.39*** | 0.097 | 48% | 89% |
| q-factor (MKT ME IA ROE) | 2.306 | 0.165 | 0.136 | 0.092 | 46% | 2.33*** | 0.103 | 51% | 87% |
| BS (MKT SMB HML$^m$ IA ROE UMD) | 2.782 | 0.199 | 0.180 | 0.083 | 21% | 2.54*** | 0.146 | 72% | 90% |
| All (all ten factors) | 2.407 | 0.172 | 0.152 | 0.080 | 28% | 2.37*** | 0.118 | 58% | 91% |
| **Panel B: 15\*10** | | | | | | | | | |
| CAPM (MKT) | 2.902 | 0.237 | 0.219 | 0.089 | 17% | 2.47*** | 0.159 | 142% | 83% |
| FF3 (MKT SMB HML) | 2.616 | 0.214 | 0.199 | 0.077 | 15% | 2.38*** | 0.115 | 102% | 87% |
| FF4 (MKT SMB HML UMD) | 1.925 | 0.157 | 0.138 | 0.075 | 29% | 2.36*** | 0.090 | 80% | 88% |
| FF5 (MKT SMB HML RMW CMA) | 2.033 | 0.166 | 0.148 | 0.075 | 26% | 2.17*** | 0.102 | 91% | 89% |
| FF6 (MKT SMB HML RMW CMA UMD) | 1.649 | 0.135 | 0.114 | 0.072 | 41% | 2.21*** | 0.086 | 76% | 89% |
| FF6-HML (MKT SMB RMW CMA UMD) | 1.652 | 0.135 | 0.112 | 0.075 | 44% | 2.20*** | 0.086 | 77% | 89% |
| q-factor (MKT ME IA ROE) | 1.858 | 0.152 | 0.128 | 0.082 | 41% | 2.34*** | 0.100 | 89% | 87% |
| BS (MKT SMB HML$^m$ IA ROE UMD) | 2.201 | 0.180 | 0.162 | 0.077 | 23% | 2.55*** | 0.133 | 119% | 89% |
| All (all ten factors) | 1.950 | 0.159 | 0.140 | 0.076 | 30% | 2.33*** | 0.112 | 100% | 90% |

Table 9. Distance-based Metrics under Varying Prior Degrees of Mispricing Uncertainty
1967:01 – 2016:12, 600 Observations

This table reports the distance-based metric *AD* and its components at varying prior degrees of mispricing uncertainty. The results for $\sigma_\alpha$ = 0, 2%, 4%, 6%, 8%, 10% are shown in Panel A – F, respectively. The first four columns are for the augmented cross section of four sets of 5×5 portfolios and three sets of 2×4×4 portfolios. The last four columns are for the augmented cross section of 15 sets of decile portfolios. For $\sigma_\alpha = 0$, $AD = \sqrt{\sum_{i=1}^n (\tilde{\alpha}_i^2 + \tilde{\sigma}_{\alpha,i}^2)/n}$ is the average distance. $RMSE(\tilde{\alpha}) = \sqrt{\sum_{i=1}^n \tilde{\alpha}_i^2/n}$ is the square root of the mean square pricing error. $RMSE(\tilde{\sigma}_\alpha) = \sqrt{\sum_{i=1}^n \tilde{\sigma}_{\alpha,i}^2/n}$ is the square root of the mean square standard error. $A\tilde{\sigma}_\alpha^2/A\tilde{\alpha}^2 = MSE(\tilde{\sigma}_\alpha)/MSE(\tilde{\alpha})$ is the ratio of the mean square standard errors to the mean square pricing error. For $\sigma_\alpha > 0$, $AD = \sqrt{(||\tilde{\alpha}_{II} - \tilde{\alpha}_I||^2 + ||\tilde{V}_{II} - \tilde{V}_I||)/n}$, where $||\tilde{\alpha}_{II} - \tilde{\alpha}_I||$ is the Euclidean 2-norm of the alpha difference vector; $||\tilde{V}_{II} - \tilde{V}_I|| = Tr(\tilde{V}_I + \tilde{V}_{II} - 2(\tilde{V}_I^{1/2}\tilde{V}_{II}\tilde{V}_I^{1/2})^{1/2})$. $RMSE(\tilde{\alpha}) = \sqrt{||\tilde{\alpha}_{II} - \tilde{\alpha}_I||^2/n}$ and $RMSE(\tilde{\sigma}_\alpha) = \sqrt{||\tilde{V}_{II} - \tilde{V}_I||/n}$. $A\tilde{\sigma}_\alpha^2/A\tilde{\alpha}^2 = ||\tilde{V}_{II} - \tilde{V}_I||/||\tilde{\alpha}_{II} - \tilde{\alpha}_I||^2$. $Tr(\cdot)$ is the trace operator of a matrix; $V_I^{1/2}$ is the square root of the alpha covariance matrix generated by model *I*. Model *I* is specified with a certain prior degree of mispricing uncertainty ($\sigma_\alpha$ = 2%, 4%, 6%, 8%, 10%), and model *II* is purely data based ($\sigma_\alpha = \infty$).

| | Panel A: $\sigma_\alpha = 0$ | | | | | | | |
|---|---|---|---|---|---|---|---|---|
| | 4*25+3*32 | | | | 15*10 | | | |
| Model (Factors) | AD | $RMSE(\tilde{\alpha})$ | $RMSE(\tilde{\sigma}_\alpha)$ | $A\tilde{\sigma}_\alpha^2/A\tilde{\alpha}^2$ | AD | $RMSE(\tilde{\alpha})$ | $RMSE(\tilde{\sigma}_\alpha)$ | $A\tilde{\sigma}_\alpha^2/A\tilde{\alpha}^2$ |
| CAPM (MKT) | 0.345 | 0.322 | 0.123 | 15% | 0.237 | 0.219 | 0.089 | 17% |
| FF3 (MKT SMB HML) | 0.241 | 0.226 | 0.083 | 13% | 0.214 | 0.199 | 0.077 | 15% |
| FF4 (MKT SMB HML UMD) | 0.185 | 0.167 | 0.080 | 23% | 0.157 | 0.138 | 0.075 | 29% |
| FF5 (MKT SMB HML RMW CMA) | 0.187 | 0.170 | 0.080 | 22% | 0.166 | 0.148 | 0.075 | 26% |
| FF6 (MKT SMB HML RMW CMA UMD) | 0.145 | 0.124 | 0.076 | 38% | 0.135 | 0.114 | 0.072 | 40% |
| FF6-HML (MKT SMB RMW CMA UMD) | 0.149 | 0.126 | 0.081 | 41% | 0.135 | 0.112 | 0.075 | 45% |
| q-factor (MKT ME IA ROE) | 0.165 | 0.136 | 0.092 | 46% | 0.152 | 0.128 | 0.082 | 41% |
| BS (MKT SMB HML$^m$ IA ROE UMD) | 0.199 | 0.180 | 0.083 | 21% | 0.180 | 0.162 | 0.077 | 23% |
| | Panel B: $\sigma_\alpha = 0.02$ | | | | | | | |
| | 4*25+3*32 | | | | 15*10 | | | |
| Model (Factors) | AD | $RMSE(\tilde{\alpha})$ | $RMSE(\tilde{\sigma}_\alpha)$ | $A\tilde{\sigma}_\alpha^2/A\tilde{\alpha}^2$ | AD | $RMSE(\tilde{\alpha})$ | $RMSE(\tilde{\sigma}_\alpha)$ | $A\tilde{\sigma}_\alpha^2/A\tilde{\alpha}^2$ |
| CAPM (MKT) | 0.330 | 0.314 | 0.103 | 11% | 0.220 | 0.209 | 0.070 | 11% |
| FF3 (MKT SMB HML) | 0.222 | 0.213 | 0.064 | 9% | 0.195 | 0.187 | 0.058 | 10% |
| FF4 (MKT SMB HML UMD) | 0.168 | 0.157 | 0.061 | 15% | 0.140 | 0.129 | 0.056 | 19% |
| FF5 (MKT SMB HML RMW CMA) | 0.171 | 0.160 | 0.060 | 14% | 0.149 | 0.138 | 0.056 | 16% |
| FF6 (MKT SMB HML RMW CMA UMD) | 0.129 | 0.116 | 0.057 | 24% | 0.118 | 0.106 | 0.053 | 25% |
| FF6-HML (MKT SMB RMW CMA UMD) | 0.133 | 0.118 | 0.062 | 28% | 0.119 | 0.105 | 0.055 | 28% |
| q-factor (MKT ME IA ROE) | 0.149 | 0.130 | 0.073 | 32% | 0.136 | 0.120 | 0.063 | 27% |
| BS (MKT SMB HML$^m$ IA ROE UMD) | 0.182 | 0.171 | 0.064 | 14% | 0.163 | 0.152 | 0.058 | 15% |

Table 9 (Cont.)

| | Panel C: $\sigma_\alpha = 0.04$ | | | | | | | |
| | 4*25+3*32 | | | | 15*10 | | | |
| Model (Factors) | AD | RMSE($\tilde{\alpha}$) | RMSE($\tilde{\sigma}_\alpha$) | $A\tilde{\sigma}_\alpha^2/A\tilde{\alpha}^2$ | AD | RMSE($\tilde{\alpha}$) | RMSE($\tilde{\sigma}_\alpha$) | $A\tilde{\sigma}_\alpha^2/A\tilde{\alpha}^2$ |
|---|---|---|---|---|---|---|---|---|
| CAPM (MKT) | 0.303 | 0.291 | 0.085 | 8% | 0.190 | 0.183 | 0.053 | 8% |
| FF3 (MKT SMB HML) | 0.189 | 0.183 | 0.047 | 7% | 0.162 | 0.157 | 0.042 | 7% |
| FF4 (MKT SMB HML UMD) | 0.141 | 0.134 | 0.045 | 11% | 0.115 | 0.108 | 0.040 | 14% |
| FF5 (MKT SMB HML RMW CMA) | 0.143 | 0.136 | 0.044 | 11% | 0.122 | 0.115 | 0.040 | 12% |
| FF6 (MKT SMB HML RMW CMA UMD) | 0.105 | 0.097 | 0.041 | 18% | 0.095 | 0.087 | 0.037 | 18% |
| FF6-HML (MKT SMB RMW CMA UMD) | 0.111 | 0.101 | 0.045 | 20% | 0.096 | 0.087 | 0.039 | 20% |
| q-factor (MKT ME IA ROE) | 0.128 | 0.115 | 0.056 | 23% | 0.113 | 0.103 | 0.046 | 20% |
| BS (MKT SMB $HML^m$ IA ROE UMD) | 0.154 | 0.147 | 0.047 | 10% | 0.135 | 0.128 | 0.042 | 11% |
| | Panel D: $\sigma_\alpha = 0.06$ | | | | | | | |
| | 4*25+3*32 | | | | 15*10 | | | |
| Model (Factors) | AD | RMSE($\tilde{\alpha}$) | RMSE($\tilde{\sigma}_\alpha$) | $A\tilde{\sigma}_\alpha^2/A\tilde{\alpha}^2$ | AD | RMSE($\tilde{\alpha}$) | RMSE($\tilde{\sigma}_\alpha$) | $A\tilde{\sigma}_\alpha^2/A\tilde{\alpha}^2$ |
| CAPM (MKT) | 0.269 | 0.260 | 0.069 | 7% | 0.156 | 0.151 | 0.039 | 7% |
| FF3 (MKT SMB HML) | 0.152 | 0.148 | 0.034 | 5% | 0.128 | 0.124 | 0.030 | 6% |
| FF4 (MKT SMB HML UMD) | 0.112 | 0.107 | 0.032 | 9% | 0.089 | 0.084 | 0.028 | 11% |
| FF5 (MKT SMB HML RMW CMA) | 0.113 | 0.108 | 0.032 | 9% | 0.095 | 0.091 | 0.028 | 10% |
| FF6 (MKT SMB HML RMW CMA UMD) | 0.082 | 0.077 | 0.029 | 14% | 0.072 | 0.067 | 0.026 | 15% |
| FF6-HML (MKT SMB RMW CMA UMD) | 0.087 | 0.081 | 0.033 | 16% | 0.074 | 0.068 | 0.028 | 17% |
| q-factor (MKT ME IA ROE) | 0.105 | 0.096 | 0.042 | 19% | 0.090 | 0.083 | 0.034 | 16% |
| BS (MKT SMB $HML^m$ IA ROE UMD) | 0.124 | 0.119 | 0.034 | 8% | 0.106 | 0.101 | 0.030 | 9% |
| | Panel E: $\sigma_\alpha = 0.08$ | | | | | | | |
| | 4*25+3*32 | | | | 15*10 | | | |
| Model (Factors) | AD | RMSE($\tilde{\alpha}$) | RMSE($\tilde{\sigma}_\alpha$) | $A\tilde{\sigma}_\alpha^2/A\tilde{\alpha}^2$ | AD | RMSE($\tilde{\alpha}$) | RMSE($\tilde{\sigma}_\alpha$) | $A\tilde{\sigma}_\alpha^2/A\tilde{\alpha}^2$ |
| CAPM (MKT) | 0.233 | 0.226 | 0.056 | 6% | 0.125 | 0.122 | 0.030 | 6% |
| FF3 (MKT SMB HML) | 0.120 | 0.117 | 0.025 | 5% | 0.098 | 0.096 | 0.022 | 5% |
| FF4 (MKT SMB HML UMD) | 0.087 | 0.084 | 0.024 | 8% | 0.068 | 0.065 | 0.020 | 10% |
| FF5 (MKT SMB HML RMW CMA) | 0.088 | 0.085 | 0.023 | 8% | 0.072 | 0.070 | 0.020 | 9% |
| FF6 (MKT SMB HML RMW CMA UMD) | 0.063 | 0.059 | 0.021 | 13% | 0.054 | 0.051 | 0.019 | 13% |
| FF6-HML (MKT SMB RMW CMA UMD) | 0.068 | 0.064 | 0.024 | 14% | 0.056 | 0.052 | 0.020 | 15% |
| q-factor (MKT ME IA ROE) | 0.084 | 0.078 | 0.032 | 17% | 0.070 | 0.065 | 0.025 | 14% |
| BS (MKT SMB $HML^m$ IA ROE UMD) | 0.097 | 0.094 | 0.025 | 7% | 0.081 | 0.078 | 0.022 | 8% |
| | Panel F: $\sigma_\alpha = 0.10$ | | | | | | | |
| | 4*25+3*32 | | | | 15*10 | | | |
| Model (Factors) | AD | RMSE($\tilde{\alpha}$) | RMSE($\tilde{\sigma}_\alpha$) | $A\tilde{\sigma}_\alpha^2/A\tilde{\alpha}^2$ | AD | RMSE($\tilde{\alpha}$) | RMSE($\tilde{\sigma}_\alpha$) | $A\tilde{\sigma}_\alpha^2/A\tilde{\alpha}^2$ |
| CAPM (MKT) | 0.199 | 0.194 | 0.045 | 5% | 0.100 | 0.097 | 0.023 | 5% |
| FF3 (MKT SMB HML) | 0.094 | 0.092 | 0.019 | 4% | 0.076 | 0.074 | 0.016 | 5% |
| FF4 (MKT SMB HML UMD) | 0.068 | 0.065 | 0.018 | 7% | 0.052 | 0.050 | 0.015 | 9% |
| FF5 (MKT SMB HML RMW CMA) | 0.068 | 0.066 | 0.017 | 7% | 0.056 | 0.054 | 0.015 | 8% |
| FF6 (MKT SMB HML RMW CMA UMD) | 0.048 | 0.046 | 0.016 | 12% | 0.041 | 0.039 | 0.014 | 12% |
| FF6-HML (MKT SMB RMW CMA UMD) | 0.053 | 0.050 | 0.018 | 13% | 0.043 | 0.040 | 0.015 | 14% |
| q-factor (MKT ME IA ROE) | 0.067 | 0.063 | 0.024 | 15% | 0.055 | 0.051 | 0.019 | 13% |
| BS (MKT SMB $HML^m$ IA ROE UMD) | 0.076 | 0.074 | 0.019 | 7% | 0.063 | 0.061 | 0.016 | 7% |